\documentclass{aa}

\usepackage[varg]{txfonts}

\usepackage{hyperref} 
\hypersetup{colorlinks=true, breaklinks = true, linkcolor=blue, filecolor=blue, urlcolor=blue, pdfpagemode=FullScreen, citecolor=blue}

\usepackage{natbib,twoopt}
\bibpunct{(}{)}{;}{a}{}{,}  

\newcommand{\It}{\Tilde{I}}

\newcommand{\Inu}{I_{\nu}}
\newcommand{\Inustar}{I_{\nu}^{\mathrm{\star}}}
\newcommand{\Inuenv}{I_{\nu}^{\mathrm{env}}}

\newcommand{\Ih}{\vec{\It}_h}
\newcommand{\Jnu}{J_{\nu}}
\newcommand{\Jnustar}{J_{\nu}^{\mathrm{\star}}}
\newcommand{\Jnuenv}{J_{\nu}^{\mathrm{env}}}
\newcommand{\mustar}{\mu_{\mathrm{\star}}}

\newcommand{\Bnu}{B_{\nu}}
\newcommand{\Rin}{R_{\mathrm{in}}}
\newcommand{\Rout}{R_{\mathrm{out}}}
\newcommand{\Rstar}{R_\mathrm{\star}}
\newcommand{\Tstar}{T_\mathrm{\star}}

\newcommand{\Tin}{T_{\mathrm{in}}}

\newcommand{\Knusca}{\kappa^{\mathrm{sca}}_{\nu}}
\newcommand{\Knuabs}{\kappa^{\mathrm{abs}}_{\nu}}
\newcommand{\Knuext}{\kappa^{\mathrm{ext}}_{\nu}}
\newcommand{\Cnusca}{C^{\mathrm{sca}}_{\nu}}
\newcommand{\Cnuabs}{C^{\mathrm{abs}}_{\nu}}
\newcommand{\Cnuext}{C^{\mathrm{ext}}_{\nu}}
\newcommand{\vecOmega}{\vec{\Omega}}
\newcommand{\xv}{\vec{x}}
\newcommand{\rv}{\vec{r}}

\newcommand{\shat}{\hat{s}}
\newcommand{\rhat}{\hat{r}}
\newcommand{\Thetahat}{\hat{\Theta}}
\newcommand{\Phihat}{\hat{\Phi}}

\newcommand{\xt}{\vec{\Tilde{x}}}
\newcommand{\rt}{\Tilde{r}}
\newcommand{\Tt}{\Tilde{\Theta}}
\newcommand{\mut}{\Tilde{\mu}}
\newcommand{\phit}{\Tilde{\varphi}}
\newcommand{\etat}{\Tilde{\eta}}

\begin{document}

\title{Discontinuous Galerkin finite element method for the continuum radiative
transfer problem inside axis-symmetric circumstellar envelopes}

\author{J. Perdigon \and M. Faurobert \and G. Niccolini}

\institute{Universit\'e de la C\^ote d'Azur, Observatoire de la C\^ote d'Azur, CNRS, Laboratoire Lagrange, Bd de l'Observatoire, CS 34229, 06304 Nice cedex 4, France \\ \email{jeremy.perdigon@oca.eu}}

\date{Received <date>/Accepted <date>}

\abstract
{The study of the continuum radiative transfer problem inside circumstellar envelopes is both a theoretical and numerical challenge, especially in the frequency-dependent and multi-dimensional case. While approximate methods are easier to handle numerically, they often fail to accurately describe the radiation field inside complex geometries. For these cases, it is necessary to directly solve the radiative transfer equation numerically.}
{We investigate the accuracy of the discontinuous Galerkin finite element method (DGFEM hereafter) applied to the frequency-dependent two-dimensional radiative transfer problem, and coupled with the radiative equilibrium equation. We next used this method in the context of axis-symmetric circumstellar envelopes.}
{The DGFEM is a variant of finite element methods. It employs discontinuous elements and flux integrals along their boundaries, ensuring local flux conservation. However, as opposed to the classical finite element methods, the solution is discontinuous across element edges. We implemented this approach in a code and tested its accuracy by comparing our results with the benchmarks from the literature.}
{For all the tested cases, the temperatures profiles agree within one percent. Additionally, the emerging spectral energy distributions (SEDs) and images, obtained by ray-tracing techniques from the DGFEM emissivity, agree on average within $5~\mathrm{\%}$ and $10~\mathrm{\%}$, respectively.}
{We show that the DGFEM can accurately describe the continuum radiative transfer problem inside axis-symmetric circumstellar envelopes. Consecutively the emerging SEDs and images are also well reproduced. The DGFEM provides an alternative method (other than Monte-Carlo methods for instance) for solving the radiative transfer equation, and it could be used in cases that are more difficult to handle with the other methods.}


\titlerunning{DGFEM for the continuum radiative transfer problem inside axis-symmetric circumstellar envelopes.}
\authorrunning{J. Perdigon et al.}

\maketitle

\section{Introduction}

The study of the continuum radiative transfer problem is crucial for the characterisation of circumstellar environments. Radiative processes play a major role in the determination of physical observables such as, for example, the temperature, abundances and velocity fields. The description of the radiation field is both a theoretical and numerical challenge, especially in the frequency-dependent and multi-dimensional case. 

Several directions have been followed to tackle this problem. One approach involves approximate methods. It is usually done by assuming a particular form for the radiation field, most of the time based on symmetry arguments. The radiative transfer equation is then recast into a presumably simpler equation. This is the case, for example, for the Eddington approximation, in which the radiation is assumed to be a correction to an isotropic field, yielding an accurate description in the optically thick regime. For this approximation, the radiation is described by a simple linear diffusion equation. More sophisticated approximations were later developed, for example the flux-limited diffusion \citep{1981ApJ...248..321L}, which is asymptotically correct to both optically thin and thick regimes. In the latter case, the radiation is described by a non-linear diffusion equation.

While approximate methods are easier to handle numerically, they often fail to describe the radiation field accurately inside complex geometries \citep{2013A&A...555A...7K}. For these cases, it is necessary to numerically solve the radiative transfer equation directly \citep[see][for a thorough review of the different methods]{2013ARA&A..51...63S}. A first approach is to approximate the transport operator with finite difference \citep[e.g][]{2003A&A...401..405S}, yielding a system of linear equations. This technique has the disadvantage of introducing spurious numerical oscillations and possible negative values for the intensity, due to the strong spatial and angular variations of radiation field. Other techniques, such as long and short characteristics \citep[e.g][]{2009A&A...501..383W}, rely on the integral form of the radiative transfer equation. They are generally harder to implement than finite difference methods, are numerically demanding, and can also exhibit pathological behaviours such as negative values \citep{1988JQSRT..39...67K}. However, both approaches allow explicit error control. Finally, Monte-Carlo methods \citep[e.g][]{2003CoPhC.150...99W} are amongst the most popular ones because they are easy and fast to implement, can handle complex geometries, and do not suffer from the same flaws as the other methods. However, they introduce noises that are inherent to their statistical nature. This class of methods additionally suffers from a reliable way of estimating the error, especially in optically thick regions where radiation is trapped inside the medium.

Aside from these techniques, finite element methods have already been used to solve the radiative transfer equation. A variant of it, the discontinuous Galerkin finite element method \citep[][DGFEM hereafter]{reed1973triangular}, makes use of discontinuous elements and flux integrals along their boundaries, ensuring local flux conservation. However, as opposed to the classical finite element methods, the reconstructed solution is discontinuous across element edges. One of the main strengths of the method is its ability to produce a high-order numerical scheme, meaning a coarse computational grid can be used to achieve a small error. This feature is particularly interesting in the context of radiative transfer, where we are often limited by computational resources. Furthermore, the solution presents little to no oscillations, even in the cases where the solution displays a non-smooth behaviour \citep{Nair2011}, which often happens in the case of radiative transfer. 

The DGFEM was successfully applied to the one-dimensional (1D) spherical transport problem, in the context of neutron transport \citep{2007JCoPh.223...67M}, or more recently in grey stellar atmospheres \citep{2016A&A...595A..90K}. Extensions to two-dimensional (2D) radiative transfer have been developed, in cylindrical coordinates, in the context of a non-local thermodynamic equilibrium (non-LTE) multilevel atom. \citet{1996ApJ...457..892D} (and references therein) used a DGFEM with spatial linear elements for the evaluation of the Eddington tensor. The Eddington tensor was then subsequently used for closing the radiation field tensor moment system of equations. In another study, \citet{CUI2005383} developed a DGFEM with a linear basis function in space and a step function in angle, for the computation of the radiative transfer in participating media.

In this study, we present the DGFEM applied to the frequency-dependent continuum radiative transfer problem, with isotropic scattering and coupled with the radiative equilibrium equation. We then used this method in the context of axis-symmetric dusty circumstellar environments. We show that this method can successfully determine the correct temperature profile, and allows for accurate images and spectral energy distributions (SEDs) to be computed by subsequent ray-tracing techniques. The paper is organised as follows: in Sect.~\ref{sect:RT_Problem} we describe the radiative transfer problem for axis-symmetric configurations. In Sect.~\ref{sect:FME_DG_RT} we present the DGFEM, and some of its numerical implementation features in Sect.~\ref{sect:numericals}. In Sect.~\ref{sect:num_tests}, we compare this method against spherically and axis-symmetric benchmarks from the literature. Finally, in Sect.~\ref{sect:conclusion}, we conclude and present some perspectives.


\section{Description of the problem} \label{sect:RT_Problem}

\begin{figure}
    \centering
    \resizebox{\hsize}{!}{\includegraphics{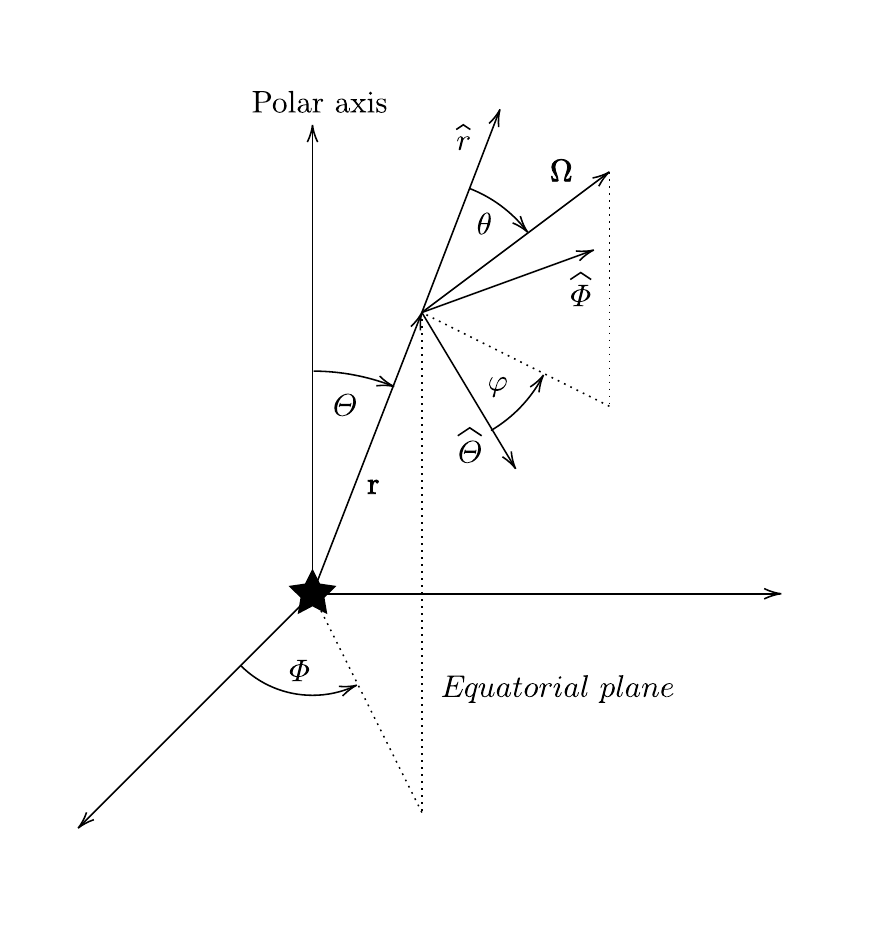}}
    \caption{Illustration showing the coordinate system. We note that $(\hat{r},\hat{\Theta},\hat{\Phi})$ is the standard spherical basis. Given the symmetry around the polar axis, the radiation field, at a given position $\vec{r}$, in a given direction $\vecOmega$ is a function of two spatial $\left(r,\Theta\right)$ and two angular $\left(\mu=\cos{\theta},\varphi\right)$ coordinates.}
    \label{fig:coordinates}
\end{figure}

\begin{figure}
    \centering
    \resizebox{\hsize}{!}{\includegraphics{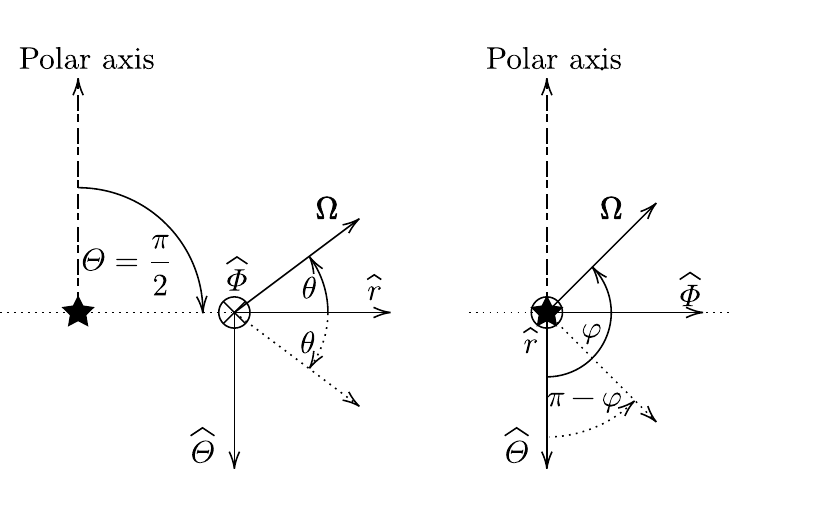}}
    \caption{Representation of the direction vector $\vecOmega$ at the equator. The dotted arrows represent the symmetric of $\vecOmega$ with respect to the equatorial plane $(\rhat,\hat{\Phi})$. The left and right 2D views allow the corresponding direction angles to be seen.}
    \label{fig:equator}
\end{figure}

We want to describe the radiation field inside an axis-symmetric circumstellar envelope. A central star of radius $\Rstar$ lies inside an inner cavity, free of matter, with a radius of $\Rin$. The envelope spans from $\Rin$ up to the outer radius $\Rout$. We assume that no matter is present after this radius. We consider the material to be exclusively made of dust. We choose to describe the problem with the spherical coordinate system, both for the spatial and angular variables (see Fig.~\ref{fig:coordinates}). In addition to the axial symmetry around the polar axis, we also assume a planar symmetry with respect to the equatorial plane, at $\Theta = \pi/2$. Given the symmetries, the domain of definition of radiation is $D \subset \mathbb{R}^4$, with $D \ni \vec{x} = \left[r, \, \Theta, \, \mu=\cos{\theta}, \, \varphi\right] \in \left[\Rin, \, \Rout \right] \times \left(0, \, \pi/2\right] \times \left[-1, \, 1\right] \times [0, \, \pi]$\footnote{Note that $\left(\rhat,\hat{\Theta}\right)$ is a plane of symmetry when axis-symmetry is assumed.}. The radiation field is described by the time-independent radiation transfer equation \citep[for a thorough derivation see e.g][II-5]{1973erh..book.....P}, 
\begin{equation}
    \begin{aligned}
        \vecOmega.\nabla \Inu + \Knuext \, \Inu = \eta_\nu, \\
        \text{with} ~ \vecOmega.\nabla \Inu = \mu \, \partial_r \, \Inu + \frac{\sqrt{1 - \mu^2} \, \cos{\varphi}}{r} \, \partial_\Theta \, \Inu + \frac{1 - \mu^2}{r} \, \partial_\mu \, \Inu \\ 
        - \frac{\cot{\Theta} \, \sqrt{1 - \mu^2} \, \sin{\varphi}}{r} \, \partial_\varphi \, \Inu. \label{eq:RTE1}
    \end{aligned} 
\end{equation}
The subscript $\nu$ denotes the frequency dependence, $\Inu$ is the specific intensity, and $\vecOmega.\nabla$ is the transport operator, which corresponds to a spatial derivative in the direction $\vecOmega$. We note that $\Knuext = \Knuabs + \Knusca$ is the extinction coefficient, with $\Knuabs$ and $\Knusca$ being the absorption and scattering coefficients, respectively. For dust species, we generally express these coefficients as $\kappa_\nu^\mathrm{abs, sca} =  C_\nu^\mathrm{abs, sca} \, n$, where $C_\nu^\mathrm{abs, sca}$ are the absorption and scattering optical cross-sections and $n$ is the number density. Dust grains are usually described as homogeneous spheres and their optical cross-sections are commonly computed with the help of Mie theory \citep{Bohren1998}. The emissivity $\eta_\nu$ includes the thermal emission (Kirchoff-Planck law), proportional to the Planck function, $\Bnu$, at the local temperature $T$, and a scattering term. For this study, we assume the scattering to be isotropic, for the purpose of our numerical tests (although it can be generalised to anisotropic scattering with no difficulty). The emissivity is then $\eta_\nu = \Knuabs \Bnu\left(T\right) + \Knusca \Jnu$, with $\Jnu$, the mean specific intensity, which is the zeroth-order angular moment of the specific intensity $\Inu$,
\begin{equation}
    \Jnu = \frac{1}{4 \, \pi} \int \limits_{4\pi} \Inu \, d\vecOmega = \frac{1}{2 \, \pi} \int \limits_{0}^{\pi} d\varphi \int \limits_{-1}^{1} \Inu \, d\mu.
\end{equation}
Additionally, the radiation field and the dust temperature are coupled via the equation of radiative equilibrium,
\begin{equation}
    4 \, \pi \int\limits_{0}^{\infty} \Knuabs \, \Bnu\left(T\right) \, d\nu = 4 \, \pi \int\limits_{0}^{\infty} \Knuabs \, \Jnu \, d\nu . \label{eq:LTE}
\end{equation}

The circumstellar matter is illuminated by a central star and it is customary to decompose the radiation field into two contributions \citep[see e.g][]{2003A&A...401..405S}, $\Inu = \Inustar + \Inuenv$, where $\Inustar$ is the direct unprocessed stellar radiation field attenuated by the circumstellar extinction, and $\Inuenv$ is the radiation emitted by the envelope (either via thermal emission or scattering). Eq.~(\ref{eq:RTE1}) can then be recast into the following system of equations,
\begin{equation}
\begin{cases}
    \vecOmega.\nabla \Inustar + \Knuext \, \Inustar = 0, \label{eq:RTE2} \\ 
    \vecOmega.\nabla \Inuenv + \Knuext \, \Inuenv = \eta_\nu =  \Knuabs \, \Bnu + \Knusca \, \left( \Jnustar + \Jnuenv \right),
\end{cases}
\end{equation}
with $\Jnustar + \Jnuenv = \Jnu$. We note that in \citet{2003A&A...401..405S}, the thermal emission term was put on the right-hand side of the first Eq.~(\ref{eq:RTE2}). However, having it in the second equation is advantageous if $\Knuabs$, $\Knusca$, and $\Knuext$ are independent of temperature. In such cases, the equations decouple and $\Inustar$ can be solved and computed definitively. The first Eq.~(\ref{eq:RTE2}) can be integrated along a given ray, yielding the formal solution for the stellar contribution,
\begin{equation}
    \Inustar =
\begin{cases}
    \Gamma_\nu^\mathrm{\star}\left(\Rstar, \, \mu\right) \, \exp{\left( -\int\limits_{0}^{s(\Rin)} \Knuext(s') \, ds'\right)},  & \text{if} ~ \mustar \leq \mu \leq 1, \\
    0, & \text{otherwise.} \label{eq:Inustar}
\end{cases}
\end{equation}
The incident stellar radiation field is $\Gamma_\nu^\mathrm{\star}\left(\Rstar, \, \mu\right)$. Again, for the purpose of the numerical tests, we assume the star to radiate as a black body at the temperature $\Tstar$. Furthermore, $\mustar = (1 - \left(\Rstar / r\right)^2)^{1/2}$ is the cosine of the angle subtended by the star at radius $r$, and $s(\Rin)$ is the distance between a given point $\vec{r}$ and $\Rin$, in the direction $\vecOmega$ . The argument in the exponential is the opposite of the optical depth integrated along the ray. For dusty media, we usually have $r \gg \Rstar$, thus the star can be treated as a point source. In the point source approximation, $\mustar \approx 1 - \left(\Rstar / r\right)^2/2$, and the stellar mean intensity can be expressed analytically as,
\begin{equation}
    \Jnustar \approx \frac{1}{4} \, \left(\frac{\Rstar}{r}\right)^2 \, \Bnu\left(\Tstar\right) \, \exp{\left( - \int \limits_{\Rin}^{r} \Knuext(r',\Theta) \, dr' \right)}. \label{eq:Jnustar}
\end{equation}
In general, the integral in Eq.~(\ref{eq:Jnustar}) can be carried out numerically, providing the stellar source term $\Jnustar$ for Eq.~(\ref{eq:RTE2}). 

To complete the description of the problem, we specify the boundary conditions for $\Inuenv$. We do so by prescribing the incident intensity $\Gamma_\nu^\mathrm{env}\left(\vec{r}_s,\vecOmega\right)$ upon the surface of the domain $D$ located at $\vec{r}_s$,
\begin{equation}
    \Inuenv\left(\vec{r}_s, \, \vecOmega\right) = \Gamma_\nu^\mathrm{env}\left(\vec{r}_s, \, \vecOmega\right), ~ \forall \, \shat.\vecOmega < 0, \label{eq:BC_RT_in}
\end{equation}
where $\shat$ is the unit vector normal to the surface of $D$, pointing outside of the domain. At the inner radius, $r=\Rin$, $\shat = -\rhat$, and the incident radiation, in a given direction $\vecOmega = \mu \, \rhat + \sqrt{1 - \mu^2} \, \cos{\varphi} \, \Thetahat + \sqrt{1 - \mu^2} \, \sin{\varphi} \, \Phihat$, comes directly from the opposite point of the cavity, 
\begin{equation}
    \Inuenv\left(\Rin, \, \Theta, \, \mu, \, \varphi\right) = \Inuenv\left(\Rin, \, \Theta', \, \mu',\, \varphi'\right), ~ \forall \, 0 < \mu < \mustar,
\end{equation}
with $(\Rin, \, \Theta', \, \mu', \, \varphi')$ being the coordinates of the opposite point. Their derivation is given in appendix \ref{appendix:bc}. On the outer edge, $r=\Rout$, $\shat = \rhat$, and we assume that there is no incident radiation upon the surface,
\begin{equation}
    \Inuenv\left(\Rout, \, \Theta, \, \mu, \, \varphi\right) = 0,  ~ \forall \,  \mu < 0.
\end{equation}
At the equator, $\Theta = \pi /2$, $\shat = \Thetahat$, and the planar symmetry requires the radiation field to be, as shown in Fig.~\ref{fig:equator},
\begin{equation}
    \Inuenv\left(r, \, \frac{\pi}{2}, \, \mu, \, \varphi\right) = \Inuenv\left(r, \, \frac{\pi}{2}, \, \mu, \, \pi-\varphi \right), ~ \forall \,  \varphi > \frac{\pi}{2}.
\end{equation}


\section{The radiative transfer equation with the discontinuous Galerkin finite element method} \label{sect:FME_DG_RT}

We present the DGFEM applied to the radiative transfer Eq.~(\ref{eq:RTE1}). We used some elements of notation from \citet{2016A&A...595A..90K} and \citet{Hesthaven2007Nodal}. For simplicity, we omitted the frequency subscript. The conservative form of Eq.~($\ref{eq:RTE1}$) is,
\begin{equation}
    \nabla_{\vec{x}}.\vec{F} + \kappa^\mathrm{ext} \, \It = \etat \, , \label{eq:RTE_conservative}
\end{equation}
with $\etat = r^2 \, \sin{\Theta} \, \eta$. We introduced the variable $\It = r^2 \, \sin{\Theta} \, I$, which is the quantity that is being conserved in Eq.~(\ref{eq:RTE_conservative}). Using this quantity is important because it improves the stability of the numerical scheme, especially near the polar axis ($\Theta = 0$) where Eq.~(\ref{eq:RTE1}) is not defined when using the spherical coordinate system. The transport operator, $\nabla_{\xv}.$, corresponds to the Cartesian divergence operator with respect to $\xv = \left( r, \, \Theta, \, \mu, \, \varphi\right)$, applied to the flux vector $\vec{F}$ of $\It$,
\begin{equation}
    \vec{F} =  \vec{a} \,  \It = \begin{pmatrix}
    a_r\\
    a_\Theta \\
    a_\mu \\
	a_ \varphi
\end{pmatrix} \, \It =  \begin{pmatrix}
   \mu \\
   \sqrt{1-\mu^2} \, \cos{\varphi} / r \\
     (1-\mu^2) / r\\
	- \cot{\Theta} \, \sqrt{1-\mu^2} \, \sin{\varphi} / r
\end{pmatrix}
\, \It. \label{eq:flux_def}
\end{equation}
We decomposed the domain $D$ into $N = N_r \times N_\Theta \times N_\mu \times N_\varphi$ non-overlapping rectangular elements $D^{i,j,k,l}$, with $N_r$, $N_\Theta$, $N_\mu$ and $N_\varphi$ being the number of elements along the $r$, $\Theta$, $\mu$ and $\varphi$ coordinates, respectively. Each element is denoted with the help of four indexes $i$, $j$ $k$, and $l$, ranging from $0$ to $N_r-1$, $N_\Theta-1$, $N_\mu-1$ and, $N_\varphi-1$. Inside each element $D^{i,j,k,l}$, we used the nodal representation and we approximated the local solution by a four-dimensional (4D) polynomial expansion,
\begin{equation}
    \It_h^{i,j,k,l}\left(\xv\right) = 
    \begin{cases}
    \sum\limits_{a=0}^{n_a-1}\sum\limits_{b=0}^{n_b-1}\sum\limits_{c=0}^{n_c-1}\sum\limits_{d=0}^{n_d-1} \It_{a,b,c,d}^{i,j,k,l} \, h_{a,b,c,d}\left(\xv\right), & \text{if} \, \xv \in D^{i,j,k,l}, \\
    0, & \text{otherwise,}
    \end{cases}
     \label{eq:Ih_def}
\end{equation}
with $n_a$, $n_b$, $n_c$, and $n_d$, being the number of nodes inside each element $D^{i,j,k,l}$, along the $r$, $\Theta$, $\mu$, and $\varphi$ coordinate, respectively. We note that $h_{a,b,c,d}$ is the 4D Lagrange polynomial, defined as,
\begin{equation}
    \begin{aligned}
     h_{a,b,c,d}\left(\xv\right) = h_a\left(r\right)  \, h_b\left(\Theta\right) \,  h_c\left(\mu\right)  \, h_d\left(\varphi\right)  \\ 
     = \prod\limits_{\substack{\alpha=0 \\ \alpha\neq a}}^{n_a-1} \frac{r-r_{\alpha}}{r_a - r_{\alpha}}  \prod\limits_{\substack{\beta=0 \\ \beta\neq b}}^{n_b-1} \frac{\Theta-\Theta_{\beta}}{\Theta_b - \Theta_{\beta}} \prod\limits_{\substack{\gamma=0 \\ \gamma\neq c}}^{n_c-1} \frac{\mu-\mu_{\gamma}}{\mu_c - \mu_{\gamma}} \prod\limits_{\substack{\delta=0 \\ \delta\neq d}}^{n_d-1} \frac{\varphi-\varphi_{\delta}}{\varphi_d - \varphi_{\delta}}. \label{eq:lagrange_poly}
    \end{aligned}
\end{equation}
By definition, the coefficients $\It_{a,b,c,d}^{i,j,k,l} = \It_h^{i,j,k,l}\left(\xv_{a,b,c,d}\right)$ correspond to the value of $\It_h^{i,j,k,l}$ at the nodes of coordinates $\xv_{a,b,c,d} = \left(r_a, \, \Theta_b, \, \mu_c, \, \varphi_d\right)$. An example of an element $D^{i,j,k,l}$ with the associated nodes is shown in Fig.~\ref{fig:element}. The global numerical approximation of the solution $I_h$ across the domain $D$ is formed by the sum of the $N$ piece-wise continuous solutions inside each element,
\begin{equation}
    \It\left(\xv\right) \approx \It_h\left(\xv\right) = \sum\limits_{i=0}^{N_r-1}\sum\limits_{j=0}^{N_\Theta-1}\sum\limits_{k=0}^{N_\mu-1} \sum\limits_{l=0}^{N_\varphi-1} \It_h^{i,j,k,l}\left(\xv\right).
\end{equation}

Now we shall introduce $\mathcal{R}_h$, the residual of Eq.~(\ref{eq:RTE_conservative}),
\begin{equation}
    \mathcal{R}_h\left(\xv\right) =  \nabla_{\vec{x}}.\vec{F}_h +  \kappa^\mathrm{ext} \, \It_h - \etat, \label{eq:residual}
\end{equation}
with $\vec{F}_h = \vec{a}\,  \It_h$. We form the classical Galerkin formulation \citep[see e.g Eq.~2.3 of][]{Hesthaven2007Nodal} by requiring the residual to be orthogonal, inside each element, to the same set of functions used for the solution representation,
\begin{equation}
    \int \limits_{D^{i,j,k,l}} \mathcal{R}_h\left(\xv\right) h_{a',b',c',d'}\left(\xv\right) \, d^4\xv = 0, ~ \forall \, D^{i,j,k,l}, \, \text{and} \,  \forall ~  h_{a',b',c',d'}. \label{eq:DG_FEM_1}
\end{equation}
The divergence term that appears in Eq.~(\ref{eq:DG_FEM_1}) can be recast with the help of the divergence theorem (Green-Ostrogradsky), yielding the following so-called weak formulation of the radiative transfer equation Eq.~(\ref{eq:RTE_conservative}):
\begin{equation}
    \begin{aligned}
    \oint \limits_{\partial D^{i,j,k,l}}^{} \shat.\vec{F}^{\mathrm{*}} \,  h_{a',b',c',d'} ~ \mathrm{d}^3\xv - \int \limits_{D^{i,j,k,l}}^{} \vec{F}_h.\nabla h_{a',b',c',d'}  ~ \mathrm{d}^4\xv \\  
    + \int \limits_{D^{i,j,k,l}}^{} \left( \kappa^\mathrm{ext} \It_h - \etat  \right) h_{a',b',c',d'} ~ \mathrm{d}^4\xv = 0,  ~  \forall \,  D^{i,j,k,l}, \, \text{and} \, \forall \, h_{a',b',c',d'}.  \label{eq:DG_FEM}
    \end{aligned}
\end{equation}
The first term, in the left-hand side of Eq.~(\ref{eq:DG_FEM}), is a surface integral, along the boundaries of $D^{i,j,k,l}$. Furthermore, $\shat$ is the outward normal vector to the surface element, and $\vec{F}^{\mathrm{*}}$ is an estimate of the flux at the cell interface, referred to as the numerical flux. It arises because the solution is not uniquely defined at the edge of the element, due to the discontinuous nature of the solution. We would like to emphasise that $D^{i,j,k,l}$ is a 4D rectangular element, and hence that each element has $2\times4$ boundary surfaces. The second and third terms in Eq.~(\ref{eq:DG_FEM}) are volume integrals and are purely local terms. Consequently, they only depend on the solution inside the element considered. For $\vec{F}^{\mathrm{*}}$, several choices are possible, depending on the nature of the problem \citep{2003ZaMM...83..731C}. In our case, and as is commonly employed for transport problems, we used the Lax-Friedrichs numerical flux, defined as (e.g on the radial right element edge where $\shat = \rhat$),
\begin{equation}
    \rhat.\vec{F}^\mathrm{*} = \frac{1}{2}\left( F_r(\It^-) + F_r(\It^+) - |a_r| \, \left( \It^+ - \It^- \right) \right), \label{eq:LF_flux}
\end{equation}
where $\It^-$ and $\It^+$ denote the left and right values of $\It$ at the element edge, respectively, and $a_r$ and $F_r$ are as defined in Eq.~(\ref{eq:flux_def}). The same form of expression holds for the other surfaces of the element.

Eq.~(\ref{eq:DG_FEM}) can be assembled into a system of $N' = N \times n$ equations with $ n = n_a \times n_b \times n_c \times n_d$, relating the $\It_{a,b,c,d}^{i,j,k,l}$ coefficients. The integrals are numerically estimated with the help of a quadrature formula. The choice of the quadrature, with the associated roots (nodes), is not unique and is usually problem-dependent \citep[for an extensive review see e.g][]{Kopriva2010On}. In general, we can put the system of Eq.~(\ref{eq:DG_FEM}) in the form of the following linear system:
\begin{equation}
    \mathcal{A} \Ih = \vec{\mathcal{B}}, \label{eq:system}
\end{equation}
with $\Ih$ being the vector of size $N'$, which contains the solution points for the full domain $D$, $\mathcal{A}$, a sparse matrix with a size of $N'\times N'$ coupling the elements of $\Ih$, and $\vec{\mathcal{B}}$ a vector of size $N'$ containing the emissivity term $\etat$.


\section{Solution strategy and numerical considerations} \label{sect:numericals}

\begin{figure}
    \centering
        \resizebox{\hsize}{!}{\includegraphics{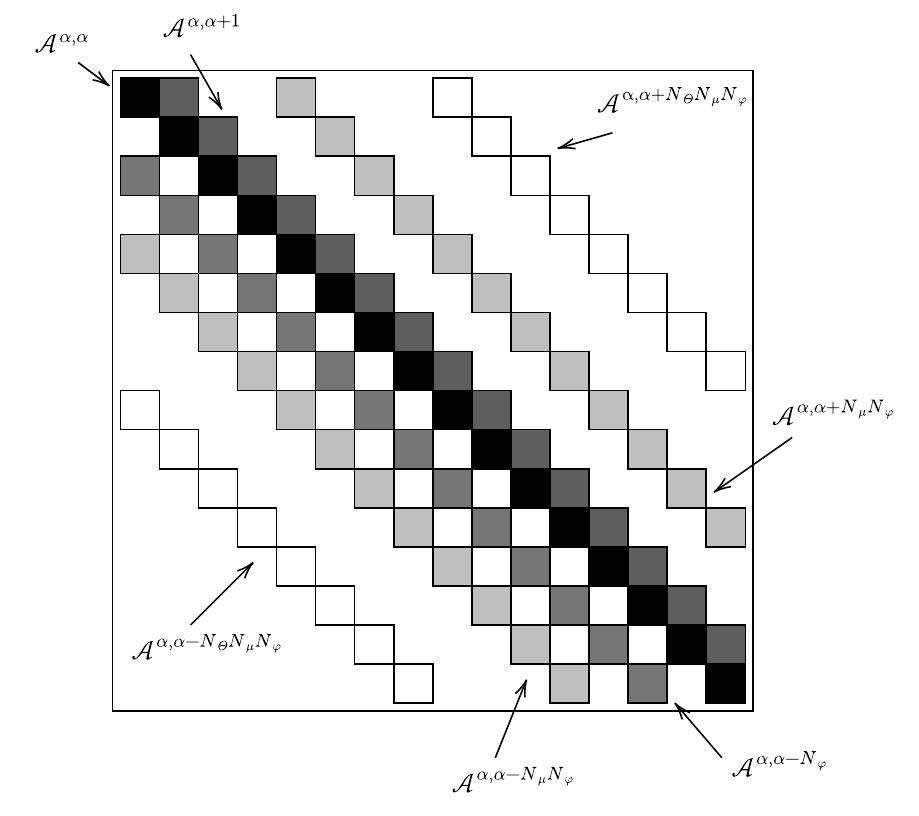}}
    \caption{Example of the sparse structure of $\mathcal{A}$ with $ N_r = N_\Theta = N_\mu = N_\varphi= 2$. The squares represent the blocks of size $n \times n$ containing non-zero values. The matrix has bands corresponding to the coupling blocks $\mathcal{A}^{\alpha,\alpha}$, $\mathcal{A}^{\alpha,\alpha+1}$, $\mathcal{A}^{\alpha,\alpha - N_\varphi}$, $\mathcal{A}^{\alpha,\alpha \pm N_\mu N_\varphi}$, and $\mathcal{A}^{\alpha,\alpha \pm N_\Theta N_\mu N_\varphi}$.}
    \label{fig:matrix_example}
\end{figure}

\begin{table*}
    \centering
    \caption{Summary of the main results}
    \begin{tabular}{c c c c c}
        \hline\hline
         & \multicolumn{2}{c}{\citet{1997MNRAS.291..121I} \, Sect.~\ref{subsect:1Dcase}}   & \multicolumn{2}{c}{\citet{2004A&A...417..793P} \, Sect.~\ref{subsect:2Dcase}} \\
        \hline
        Optical depth $\tau_{\nu_0}$ & $1$ & $10^{2}$ & $10^{-1}$ & $10^2$ \\
        $\epsilon(T)~(\mathrm{\%})$ & 0/0/0 & 0/0/1  & 0/0/0  & 1/0/2   \\
        $\epsilon\left(\text{SED}\right)_{i=12.5 \mathrm{deg}}~(\mathrm{\%})$ & -/-/- & -/-/- & 2/2/6  & 2/3/10   \\
        $\epsilon\left(\text{SED}\right)_{i=77.5 \mathrm{deg}}~(\mathrm{\%})$ & 2/2/8 & 1/2/2 & 2/2/5  & 3/4/24  \\
        \hline
    \end{tabular}
    \tablefoot{Relative differences for the temperature profiles, $\epsilon(T)$, and for the SEDs, $\epsilon\left(\text{SED}\right)$, of the two test problems. The results are presented in the form mean($|\epsilon|$) / std($|\epsilon|$) / max($|\epsilon|$) and rounded to the closest percent.}
    \label{tab:recap_tab}
\end{table*}

The solution of the problem presented in Sect.~\ref{sect:RT_Problem} involves determining the radiation field, given by Eq.~(\ref{eq:RTE2}), coupled with the equation of radiative equilibrium, Eq.~(\ref{eq:LTE}). For the test cases we consider in Sect.~\ref{sect:num_tests}, the star was treated in the point source approximation, and hence we directly used Eq.~(\ref{eq:Jnustar}) to compute $\Jnustar$. The radiation field of the envelope $\Inuenv$, described by the second Eq.~(\ref{eq:RTE2}), was solved with the DGFEM presented in Sect.~\ref{sect:FME_DG_RT}. For simplicity, we omitted the envelope superscript but implicitly refer to this contribution below.

Because the radiative transfer problem is intrinsically highly dimensional, the size of the matrix $\mathcal{A}$ is often huge and solving Eq.~(\ref{eq:system}) becomes numerically tedious. However, $\mathcal{A}$ is sparse, because each solution point, $I_{a,b,c,d}^{i,j,k,l}$, inside a given element, $D^{i,j,kl}$, only depends on the other points inside the same element and the neighbouring ones, inside $D^{i\pm1,j\pm1,k-1,l+1}$. This property allows us to rewrite Eq.~(\ref{eq:system}) as,
\begin{equation}
\begin{aligned}
    \mathcal{A}^{i,j,k,l} \, \Ih^{i,j,k,l} + \mathcal{A}^{i+1,j,k,l} \, \Ih^{i+1,j,k,l} + \mathcal{A}^{i-1,j,k,l} \, \Ih^{i-1,j,k,l} \\
    + \mathcal{A}^{i,j+1,k,l} \, \Ih^{i,j+1,k,l} + \mathcal{A}^{i,j-1,k,l} \, \Ih^{i,j-1,k,l} + \mathcal{A}^{i,j,k-1,l} \, \Ih^{i,j,k-1,l} \\
    + \mathcal{A}^{i,j,k,l+1} \, \Ih^{i,j,k,l+1} = \vec{\mathcal{B}}^{i,j,k,l}, ~ \forall \, i,\, j, \, k, \, \text{and} \, l. \label{eq:system_blocs}
\end{aligned}
\end{equation}
We note that $\mathcal{A}^{i,j,k,l}$ are the diagonal blocks of $\mathcal{A}$, with a size of $n \times n$ while $\mathcal{A}^{i\pm1,j\pm1,k-1,l+1}$ (with a size of $n \times n$) are the only non-zero, non-diagonal blocks. Furthermore, the elements $D^{i,j,k+1,l}$ and $D^{i,j,k,l-1}$ do not contribute because of the expression of the Lax-Friedrichs numerical flux, Eq.~(\ref{eq:LF_flux}), with $a_\mu \geq 0$ and $a_\varphi \leq 0$, $\forall x \in D $. We note that $\Ih^{i,j,k,l}$ and $\vec{\mathcal{B}}^{i,j,k,l}$ are the sub-vector of $\Ih$ and $\vec{\mathcal{B}}$, respectively, with a length of $n$, containing the local points in $D^{i,j,k,l}$.

To put $\mathcal{A}$ in a matrix form, we used a global index $\alpha = i \, N_\Theta \, N_\mu \, N_\varphi + j \, N_\mu \, N_\varphi + k \, N_\varphi + l$. An example of the structure of the matrix $\mathcal{A}$ with the associated blocks is displayed in Fig.~\ref{fig:matrix_example}. The formulation Eq.~(\ref{eq:system_blocs}) avoids the storage and computation of the entire matrix $\mathcal{A}$, which reduces the computational effort. 

In general, the simplest approach to obtain a solution from the problem is to solve Eq.~(\ref{eq:system_blocs}) and to update the temperature with Eq.~(\ref{eq:LTE}). Iterating between these two steps until convergence yields the solution to the problem. This procedure, commonly referred to as the $\Lambda$-iteration in the literature, becomes very slow and does not converge for large optical depths \citep[][VI-83]{1999DoverM}. Our solution strategy is directly inspired from \citet{2021A&A...653A.139P}. The key point of the method is to solve simultaneously instead of repetitively, Eq.~(\ref{eq:system_blocs}) and Eq.~(\ref{eq:LTE}). This strategy can be assimilated to an acceleration procedure to the usual $\Lambda$-iteration and yields satisfying results up to moderately thick envelopes. We however note that, as for the usual $\Lambda$-iteration, it converges very slowly for optically thick envelopes.

We proceeded with the following solution strategy: if we denote the iteration of the method by the superscript index $n$, we first (i) computed the stellar mean radiation field $\Jnustar$ with Eq.~(\ref{eq:Jnustar}) and set $\left[\Inuenv\right]^{n=0}= 0$, as an initial condition. This allowed us to compute the initial temperature profile $T^{0}$, with the help of Eq.~(\ref{eq:LTE}), where $\Jnu^0 = \Jnustar$. Then (ii), for each frequency, we computed $[\Inuenv]^{n+1}$ with the help of Eq.~(\ref{eq:system_blocs}), which we rewrote, performing a block Gauss-Seidel sweep \citep[see e.g][VII-2]{karniadakis2003},
\begin{equation}
\begin{aligned}
    \mathcal{A}^{i,j,k,l}\left[\Ih^{i,j,k,l}\right]^{n+1} =  \left[\vec{\mathcal{B}}^{i,j,k,l}\right]^{n}  -  \mathcal{A}^{i+1,j,k,l}\left[\Ih^{i+1,j,k,l}\right]^{n} \\
    - \mathcal{A}^{i-1,j,k,l}\left[\Ih^{i-1,j,k,l} \right]^{n+1}
    - \mathcal{A}^{i,j+1,k,l}\left[\Ih^{i,j+1,k,l}\right]^{n} \\
    - \mathcal{A}^{i,j-1,k,l}\left[\Ih^{i,j-1,k,l}\right]^{n+1} 
    - \mathcal{A}^{i,j,k-1,l}\left[\Ih^{i,j,k-1,l}\right]^{n+1} \\
    - \mathcal{A}^{i,j,k,l+1}\left[\Ih^{i,j,k,l+1}\right]^{n}. \label{eq:system_iteration}
\end{aligned}
\end{equation}
We give in appendix \ref{appendix:computations} the expressions of $\mathcal{A}^{i,j,k,l}$ and of the right-hand side of Eq.~(\ref{eq:system_iteration}). We note that Eq.~(\ref{eq:system_iteration}) represents $N$ linear systems to solve. We solved the linear system Eq.~(\ref{eq:system_iteration}) by direct inversion, using the Gauss elimination algorithm \citep[see e.g][IX-1]{karniadakis2003}. (iii) We computed $\Jnuenv$ and consequently updated the temperature via Eq.~(\ref{eq:LTE}). The new temperature allowed us to update the right-hand side of Eq.~(\ref{eq:system_iteration}) and to repeat steps (ii) and (iii), until convergence. We would like to point out that this scheme is not strictly identical to the usual $\Lambda$-iteration as we updated the temperature together with $\Inu$ after each iteration index $n$. Finally, concerning the implementation of the boundary conditions, we directly followed the prescription from \citet[][Appendix A.6]{2016A&A...595A..90K}.


\section{Numerical tests} \label{sect:num_tests}

 The lack of analytic solutions for the radiative transfer problem, especially in the multi-dimensional and frequency-dependent case, limited our options for testing the validity of the method. In general, one has to compare numerical solutions with other ones from the literature. For this purpose, benchmark problems have been proposed. The first test case we consider, in Sect.~\ref{subsect:1Dcase}, is the frequency-dependent radiative transfer problem inside a spherically symmetric envelope, from \citet{1997MNRAS.291..121I}. The second test case, in Sect.~\ref{subsect:2Dcase}, is about the frequency-dependent radiative transfer inside an axis-symmetric envelope  (disc), from \citet{2004A&A...417..793P}. For the latter test, we note that five different codes were used (including Monte Carlo and grid-based methods) to produce the benchmark and check the consistency of the results. We note that both of these tests are compatible with the boundary conditions presented in Sect.~\ref{sect:RT_Problem}. A summary of the main results is presented in Table \ref{tab:recap_tab}.

\subsection{1D spherically symmetric envelope} \label{subsect:1Dcase}

\begin{figure*}
    \centering
    \includegraphics[width=0.8\textwidth]{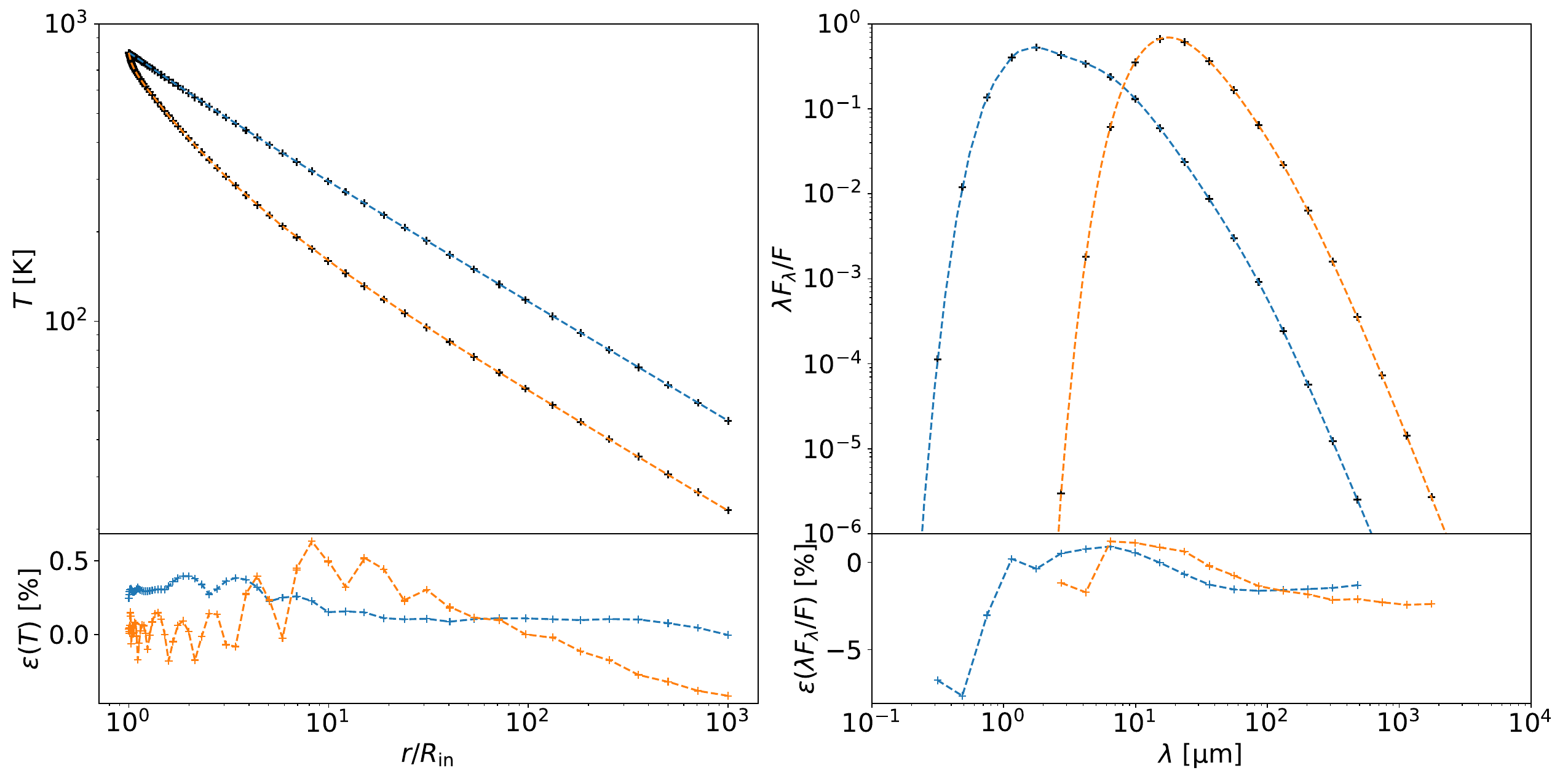}
    \caption{Temperature profiles (left panels) and normalised SEDs (right panels) for the spherically symmetric envelope, with $\tau_{\nu_0} = 1$ (blue curve) and $\tau_{\nu_0} = 10^2$ (orange curve). The cross marks represent the solution from this study and the dashed curves are from DUSTY \citep{1997MNRAS.291..121I}. The lower panels show the relative differences between the two codes.}
    \label{fig:ivezic_bench}
\end{figure*}

A point source, surrounded by a spherically symmetric envelope of dust, at radiative equilibrium, radiates as a black body at the temperature $\Tstar = 2 \, 500 ~ \mathrm{K}$. This envelope extends from the inner radius $\Rin$ to the outer radius $\Rout = 1 \, 000 ~ \Rin$. The inner radius was set so that its temperature would always be $\Tin = T(\Rin)= 800 ~ \mathrm{K}$. The number density profile, $n(r)$, is assumed to be a power law, of the form $n(r) = n_0 \left(\Rin/r\right)^{2}$. The number density at the base of the disc, $n_0$, is determined by setting the value of $\tau_{\nu_0}$ , which corresponds to the radial optical depth integrated though the envelope, at $\nu_0 = c / \lambda_0$, with $\lambda_0 = 1~\mathrm{\mu m}$,
\begin{equation}
    \tau_{\nu_0} = \int \limits_{\Rin}^{\Rout} \kappa_{\nu_0}^\mathrm{ext} \, dr = C_{\nu_0}^{\mathrm{ext}} \int \limits_{\Rin}^{\Rout} n \, dr = C_{\nu_0}^{\mathrm{ext}} \, n_0 \, \Rin \, \left( 1 - \frac{\Rin}{\Rout} \right).
\end{equation}
We note that $C_{\nu_0}^{\mathrm{ext}} = C_{\nu_0}^{\mathrm{abs}} + C_{\nu_0}^{\mathrm{sca}}$ is the extinction cross-section coefficient at $\nu_0$. In this test we consider $\tau_{\nu_0}=1$ and $10^2$ to correspond to a moderately thin and thick envelopes, respectively. The absorption and scattering cross-sections, $\Cnuabs$ and $\Cnusca$, feature a bi-linear behaviour in log-log scaling, with a constant profile for $\nu \geq \nu_0$ and a power-law dependence $ \propto \left( \nu / \nu_0 \right)^{\alpha}$, for $\nu \leq \nu_0$, with $\alpha = 1$ and $4$ for absorption and scattering, respectively. This dependence aims to mimic the behaviour of spherical astronomical silicate grains. We set the value of the thermal coupling parameter $\Cnuabs / \Cnuext$ to be $1/2$. The benchmarks from \citet{1997MNRAS.291..121I} were reproduced with version 2 of DUSTY\footnote{available at \url{http://faculty.washington.edu/ivezic/dusty_web/}}. 
Although the envelope is spherically symmetric, we used a grid of $N = N_r \times N_\Theta \times N_\mu \times N_\varphi = 16^4$ elements in our DGFEM code, with each elements containing $ n_a \times n_b \times n_c \times n_d = 3 \times 2 \times 3 \times 3$ nodes, as pictured in Fig.~\ref{fig:element}. For the radial coordinate, the cell edges were logarithmically spaced, to account for the important dynamic of the solution with respect to the radius. The frequency grid consists of $60$ logaritmically spaced points, ranging from $\lambda = 10^{-2}~\mathrm{\mu m}$ to $\lambda = 3.6 \times 10^{4}~\mathrm{\mu m}$.

The temperature profiles, $T$, and the SEDs, $\lambda F_\lambda / F$ with  $F = \int_0^\infty F_\lambda \, d\lambda$, of the envelope are shown in Fig.~\ref{fig:ivezic_bench}. For the DGFEM code, the temperature that is shown corresponds to the mean radial profile across all angular points $\Theta$. Additionally, the normalised SEDs were computed with the help of a ray-tracing module we present in Appendix \ref{appendix:rayTracer}. We subsequently explain the reasons for such a choice in Sect.~\ref{sect:grid_resolution}. The spatial and frequency grids differ between both codes and we performed a linear interpolation (in log-log scaling) of the DUSTY profiles at our grid points, in order to do the comparison. We observed a good agreement between the two codes. On average, the absolute relative differences stay below $1~\%$ for the temperatures and $2~\%$ for the SEDs. 

\subsection{2D axis-symmetric envelope} \label{subsect:2Dcase}

\begin{figure*}
    \centering
    \includegraphics[width=0.8\textwidth]{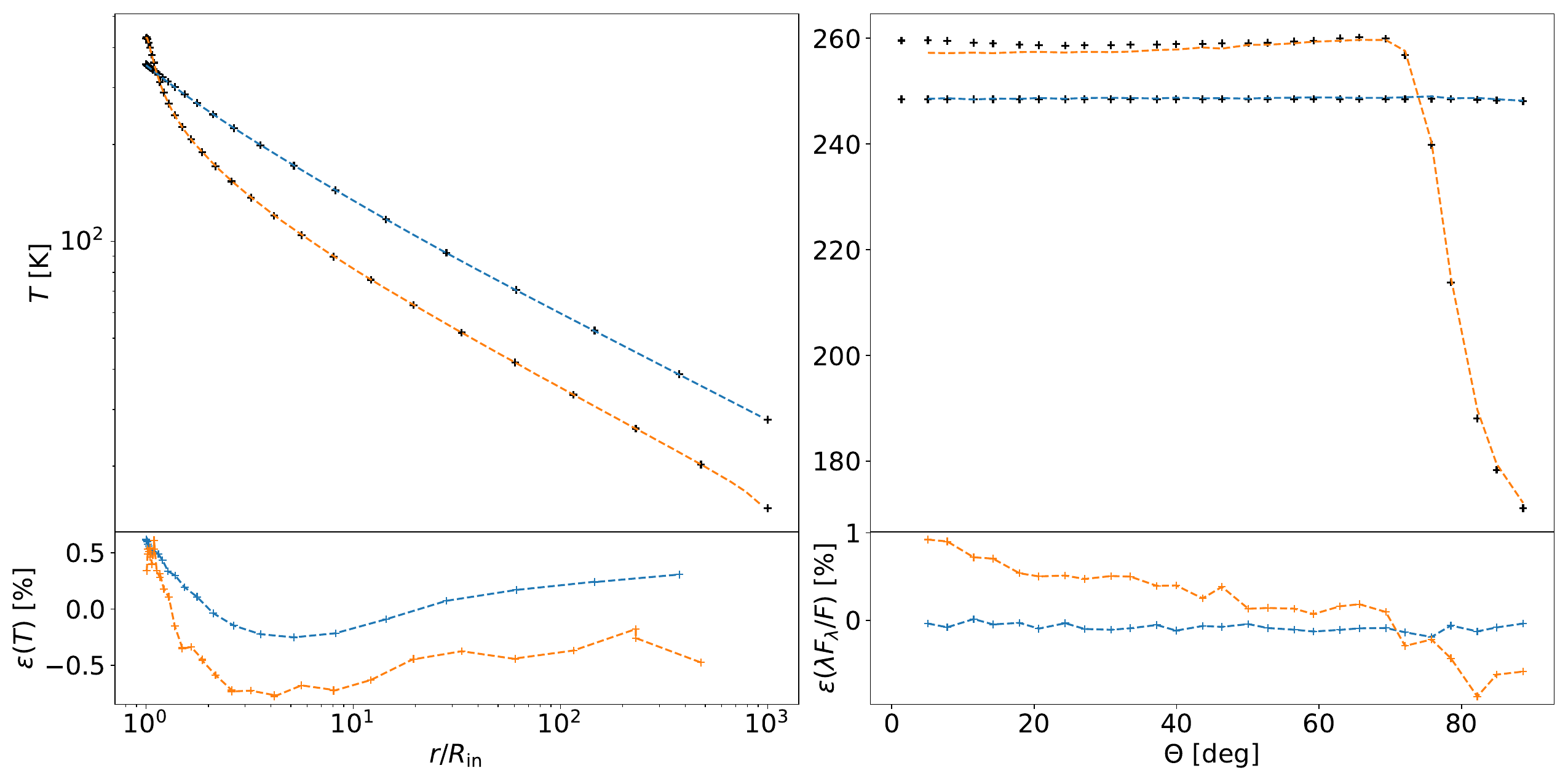}
    \caption{Temperature profiles for the axis-symmetric envelope with $\tau_{\nu_0} = 10^{-1}$ (blue curve) and $\tau_{\nu_0} = 10^2$ (orange curve), in the in the disc mid-plane (left panel) and at r=2 AU (right panel). The cross marks represent the solution from this study and the dashed curves were computed with RADMC-3D. The lower panels show the relative differences between the two codes.}
    \label{fig:pascucci_bench_temp}
\end{figure*}

\begin{figure*}
    \centering
    \includegraphics[width=0.8\textwidth]{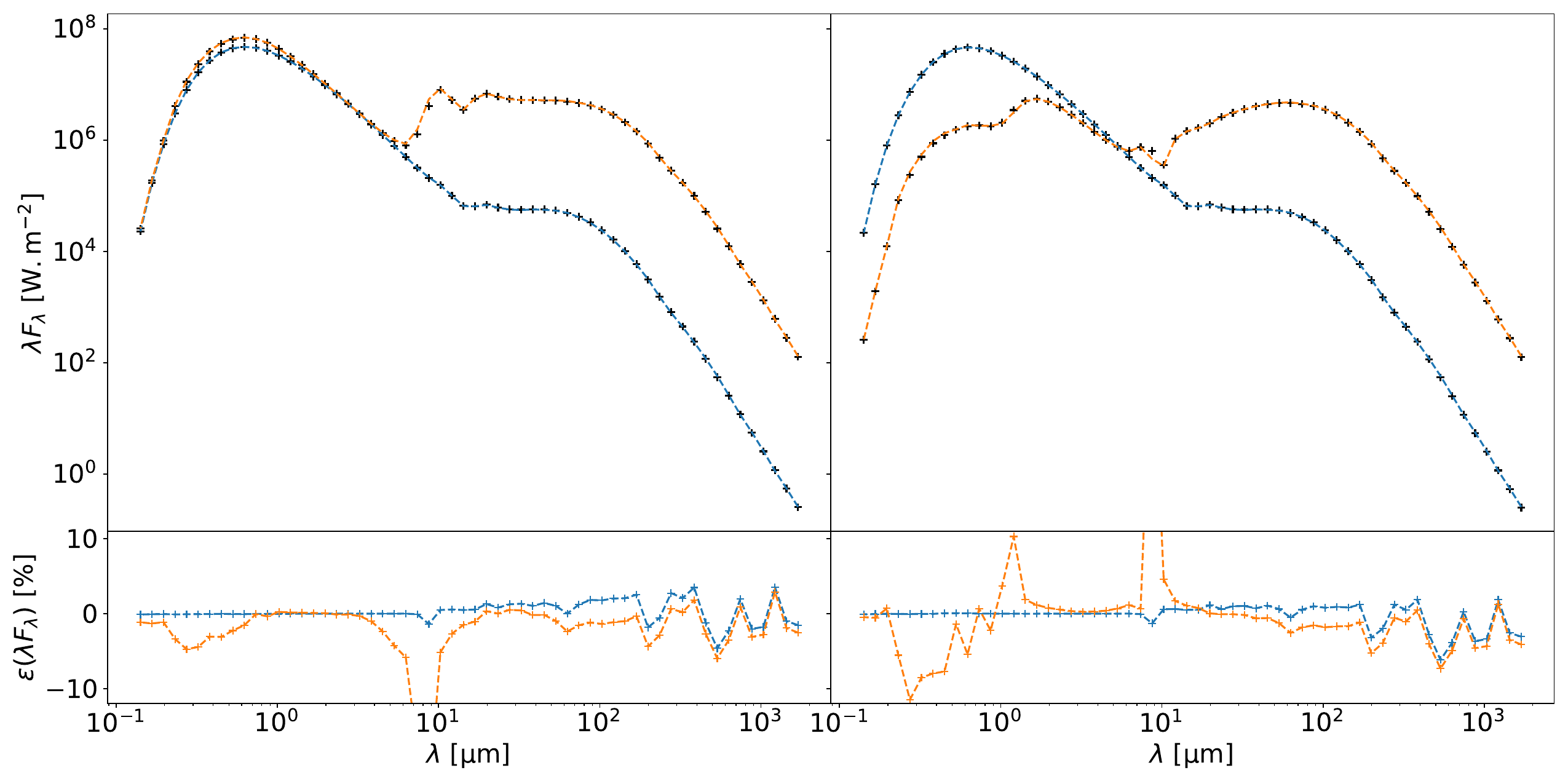}
    \caption{SED profiles for the axis-symmetric envelope with $\tau_{\nu_0} = 10^{-1}$ (blue curve) and $\tau_{\nu_0} = 10^2$ (orange curve). The left and right panels correspond to $i=12.5$ and $77.5~\mathrm{deg}$, respectively. The cross marks represent the solution from this study and the dashed curves were computed with RADMC-3D. The lower panels show the relative differences between the two codes.}
    \label{fig:pascucci_bench_sed}
\end{figure*}

\begin{figure*}
    \centering
    \includegraphics[width=0.8\textwidth]{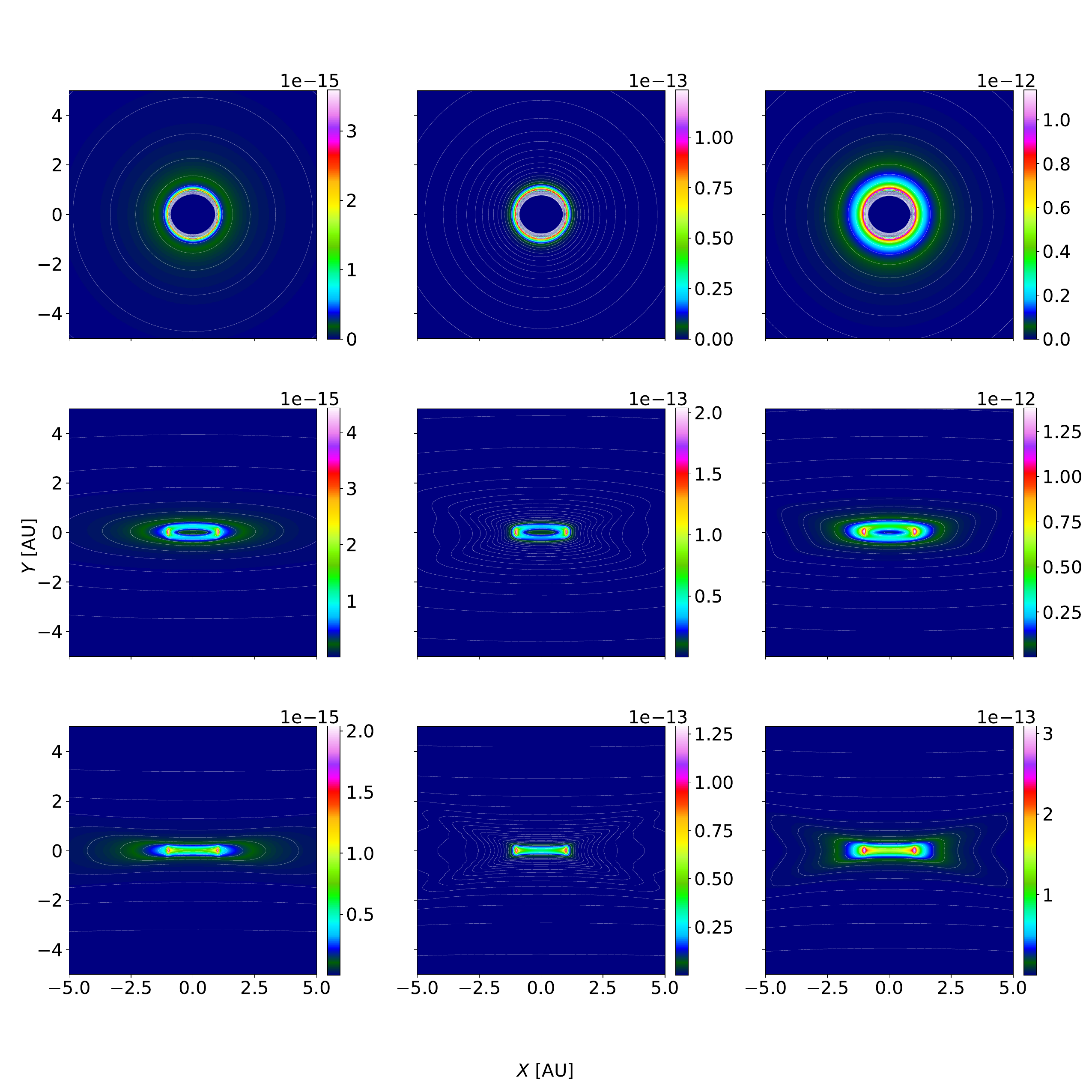}
    \caption{Images at $\lambda = 2.3, \, 4.5$, and $12.1~\mathrm{\mu m}$ (left, middle, and right panels, respectively) of the $10~\mathrm{AU}$ inner regions of the axis-symmetric envelope (Sect.~\ref{subsect:2Dcase}) and computed from the DGFEM solution with the ray-tracing module (appendix \ref{appendix:rayTracer}). The top, middle and bottom panels correspond to the inclinations $i= 12.5, 77.5,$ and $90~\mathrm{deg}$, respectively. The colour code shows the specific intensity value (in  $\mathrm{W.m^{-2}.Hz^{-1}.sr^{-1}}$) of the envelope $\Inuenv$, inside each pixel. The solid white lines show the iso-contours of $\Inuenv$.}
    \label{fig:pascucci_bench_imag}
\end{figure*}

\begin{figure*}
    \centering
    \includegraphics[width=0.8\textwidth]{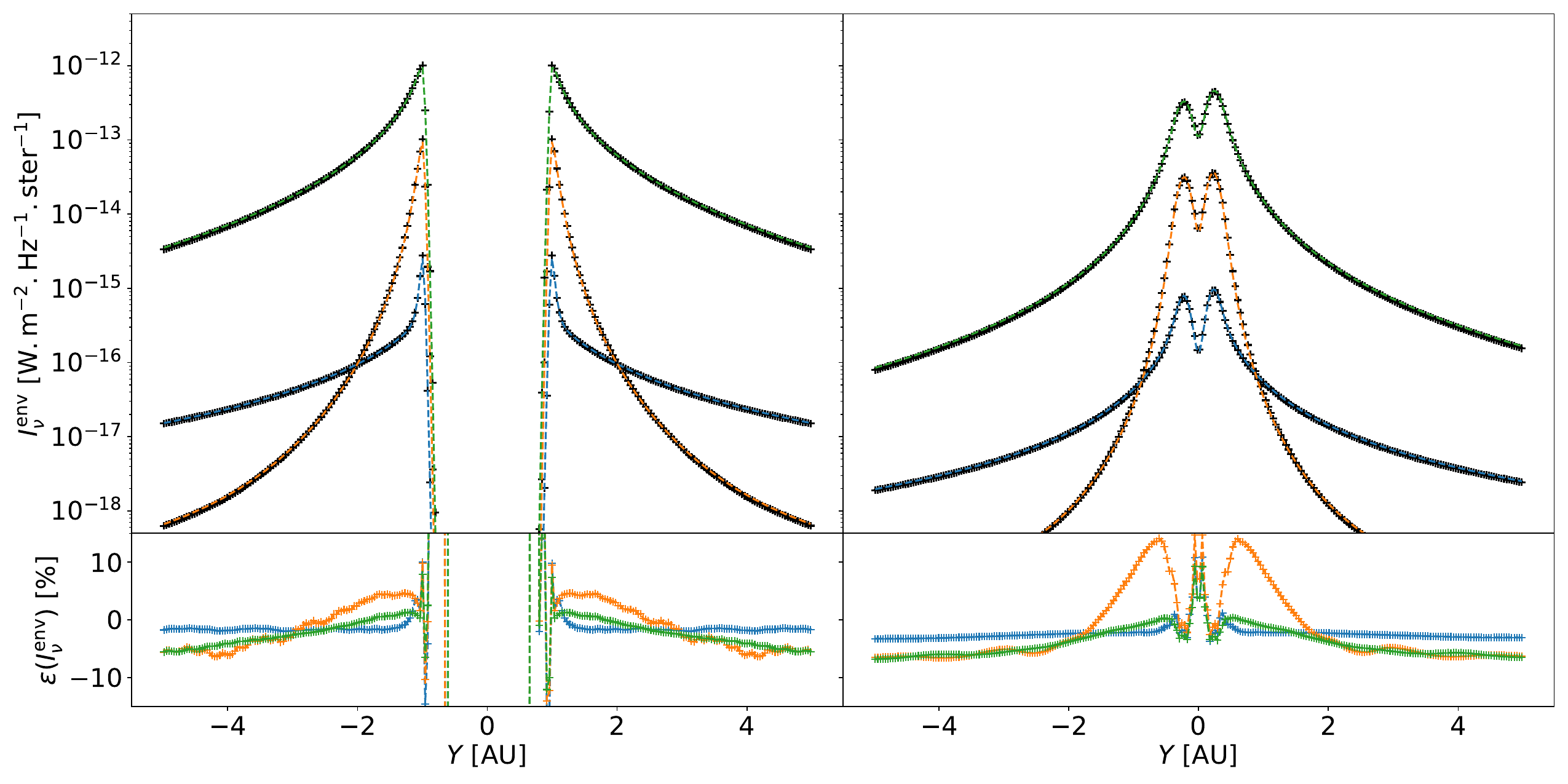}
    \caption{Image slices at $x= 0~\mathrm{AU}$, for two inclinations $i= 12.5$ and $77.5 ~ \mathrm{deg}$ (left and right panels, respectively). The blue, orange, and green curves are the RADMC-3D intensities at $\lambda = 2.3, 4.5,$ and $12.1~\mathrm{\mu m}$, respectively. The cross marks represent the solution from this study. The lower panels show the relative differences between the two codes.}
    \label{fig:pascucci_bench_imag_comp}
\end{figure*}

A point source surrounded by an axis-symmetric disc of dust, at radiative equilibrium, radiates as a black body at the temperature $\Tstar = 5 \, 800 ~ \mathrm{K}$. This disc extends from the inner radius $\Rin = 1~\mathrm{AU}$ to the outer radius $\Rout = 1 \, 000 ~ \mathrm{AU}$. The density profile $n(r,\Theta)$ is assumed to be
\begin{equation}
n(r,\Theta) = n_0 \frac{r_d}{r \, \sin{\Theta}} \exp{\left( -\frac{\pi}{4} \left(\frac{r \cos{\Theta}}{h}\right)^2 \right)},
\end{equation}
with $h = z_d \left( r \, \sin{\Theta}/r_d\right)^{\frac{9}{8}}$, $r_d = \Rout /2$, and $z_d = \Rout / 8$. This density law is characteristic of a Keplerian disc hydrostatically supported in the vertical direction, assuming a constant temperature along this direction \citep{1997ApJ...490..368C}. Again, $n_0$ was determined to set $\tau_{\nu_0}$, the radial optical depth, integrated through the disc mid-plane $\left(\Theta = \pi/2\right)$,  at frequency $\nu_0 = c/ \lambda_0$ with $\lambda_0 = 0.55~\mathrm{\mu m}$,
\begin{equation}
    n_0 = \frac{\tau_{\nu_0}}{ C_{\nu_0}^{\mathrm{ext}} \, r_d \, \ln{\left(\frac{\Rout}{\Rin}\right)}}.
\end{equation}
The opacities were taken from \citet{1984ApJ...285...89D}. They are associated with a unique dust species composed of spherical astronomical silicate grains, with a radius of $0.12 \, \mathrm{\mu m}$ and a density of $3.6 \, \mathrm{g.cm^{-3}}$. A table of pre-computed values for $\Cnuext$ and $\Cnusca$ is available in \citet{2004A&A...417..793P}. We considered the optically thin and thick cases $\tau_{\nu_0} = 10^{-1}$ and $10^2$, respectively. The benchmarks were produced with RADMC-3D\footnote{available at \url{https://www.ita.uni-heidelberg.de/~dullemond/software/radmc-3d/}} \citep{2012ascl.soft02015D}, a Monte-Carlo radiative transfer code and whose previous version is part of the original benchmarks from \citet{2004A&A...417..793P}. For our code, we used the same spatial and frequency grids as in the previous test. For RADMC-3D, we used $128$ points both in radial and angular direction, with a logarithmic sampling in the radial direction.

The temperature of the disc is displayed in Fig.~\ref{fig:pascucci_bench_temp}. The DGFEM code successfully reproduces it. The regions of the disc with steep gradients are always well reproduced, even with a fairly reasonable number of nodes. This result is a direct consequence of having a high-order numerical scheme. The radiation field can exhibit discontinuities, because of boundary conditions or very strong density gradients. Our numerical tests revealed little to no oscillations and very few negative values for the specific intensity, which did not pollute the computation of the mean radiation field. On average, the absolute relative differences stay below one percent for both test cases. The temperature in the disc mid-plane is very well reproduced, highlighting that the method is able to correctly represent the shadowed regions of the disc (the cold outer mid-plane regions shadowed by the inner dense regions). 

In Fig.~\ref{fig:pascucci_bench_sed}, we show the corresponding emerging SEDs, for two inclinations with respect to the polar axis $i=12.5$ and $77.5~\mathrm{deg}$. Again our SEDs were computed from the ray-tracing of the emissivity $\eta_\nu$. They consist of a stellar and an envelope contribution (see Eq.~\ref{eq:sum_flux}). The stellar component that peaks at around $\lambda \approx 0.6~\mathrm{\mu m}$ was computed from of Eq.~\ref{eq:fobs_star} and the discrepancies are always $<1~\mathrm{\%}$ where this part dominates. Concerning the contribution from the envelope, both the scattering ($< 1~\mathrm{\mu m}$) and emission ($> 10~\mathrm{\mu m}$) parts agree well. On average, the absolute relative differences stay below $5~\mathrm{\%}$, which are in the typical discrepancy levels of the original benchmark. We note a peak in the discrepancies for the optically thick case, between $8.7$ and $10.2~\mathrm{\mu m}$ (according to the resolution of our frequency grid). The same behaviour was previously observed in \citet{2004A&A...417..793P}, between a grid-based and a Monte-Carlo code. The authors suggested that, at these wavelengths, the flux mainly comes from the inner disc regions (between 1 and 2 AU) and the numerical simulations are particularly sensitive to the resolution of the inner parts. However, we tried to increase the resolution in these regions, which did not result in any improvement.

In Fig.~\ref{fig:pascucci_bench_imag}, we show a set of images from the DGFEM code of the $10~\mathrm{AU}$ disc inner regions, at $\lambda = 2.3, \, 4.5,$ and $12.1~\mathrm{\mu m}$ and for several inclinations $i= 12.5, \, 77.5,$ and $90~\mathrm{deg}$. These wavelengths are characteristic of the operating spectral bands of the interferometric instruments, such as GRAVITY \citep{2017A&A...602A..94G} and MATISSE \citep{2022A&A...659A.192L}. On average, the agreement between the images from the DGFEM code and the images from RADMC-3D is around $10~\mathrm{\%}$, for all frequencies and inclinations. We show in Fig.~\ref{fig:pascucci_bench_imag_comp} the comparison of an image slice at $x= 0~\mathrm{AU}$, for two inclinations $i= 12.5$ and $77.5 ~ \mathrm{deg}$ and for $\lambda = 2.3, \, 4.5,$ and $12.1~\mathrm{\mu m}$. In general, the disc emitting inner regions (peaks in Fig.~\ref{fig:pascucci_bench_imag}) are reproduced very well ($\epsilon(\Inuenv) \leq 3~\mathrm{\%}$). The biggest discrepancies occur in the wings of the peaks, where the gradient of intensity is the steepest (in logarithmic scaling).

\subsection{Effect of the grid resolution on temperatures and SEDs} \label{sect:grid_resolution}

\begin{figure}
    \centering
        \resizebox{\hsize}{!}{\includegraphics{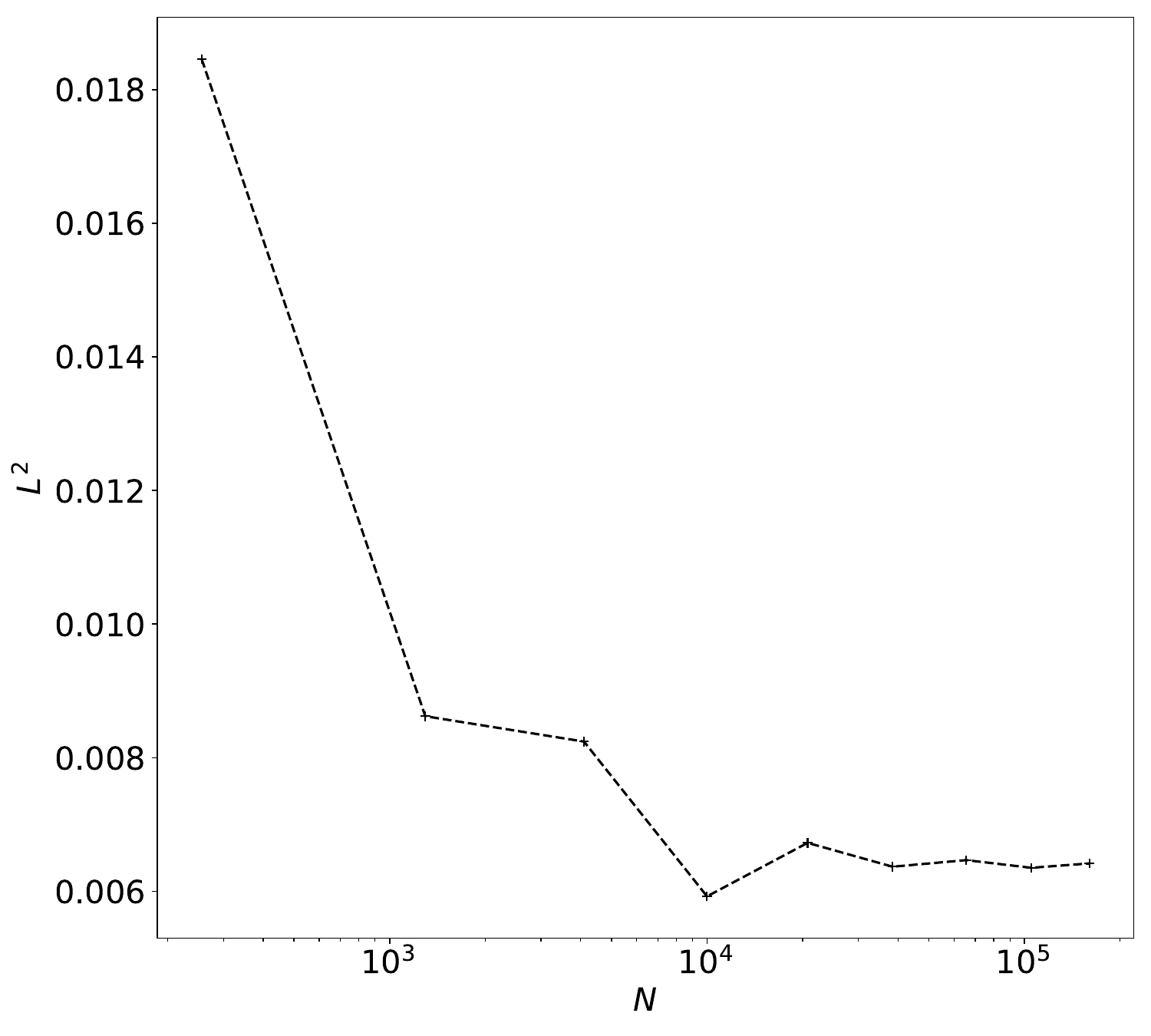}}
    \caption{$L^2$ norm of the relative differences in temperature between this study and RADMC-3D versus the number of elements in the computational grid ( $N = N_r \times N_\Theta \times N_\mu \times N_\varphi$). The $L^2$ norm was computed for the benchmark problem presented in Sect.~\ref{subsect:2Dcase}.}
    \label{fig:error_temperature}
\end{figure}

\begin{figure*}
    \centering
    \includegraphics[width=0.8\textwidth]{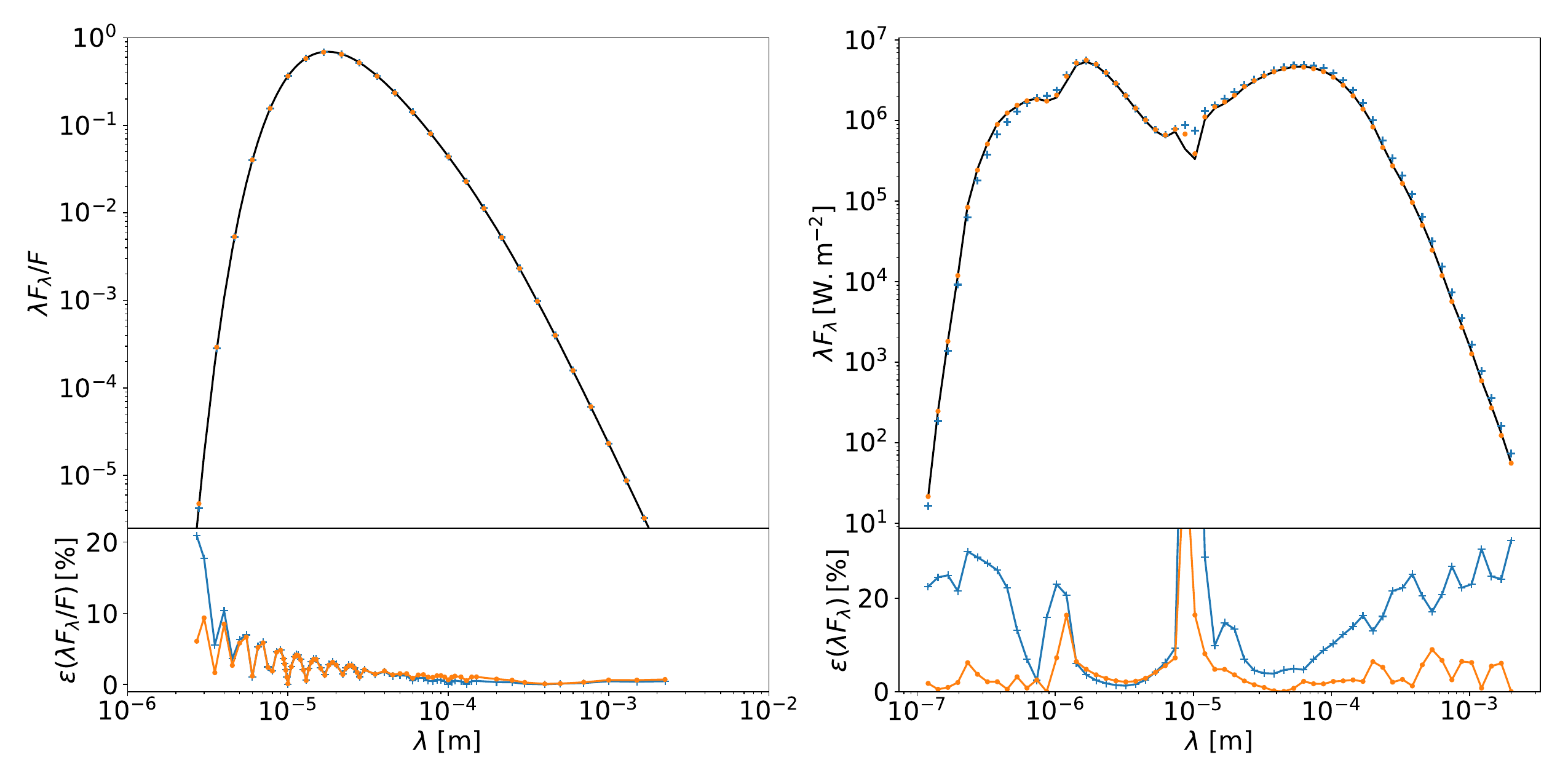}
    \caption{Emerging SEDs of RADMC-3D (solid black lines) and of the DGFEM code, computed from the post-processing ray-tracing procedure (orange dots), and from the DGFEM solution at the outer edge (blue crosses). The left panel corresponds to the optically thick spherically symmetric case ($\tau =100$, see Sect.~ \ref{subsect:1Dcase}) while the right panel corresponds to the optically thick disc benchmark from Sect.~\ref{subsect:2Dcase}. The lower panels show the relative difference between the curves of this study and the benchmarks.}
    \label{fig:comp_sed_iout_rt}
\end{figure*}

As we previously mentioned in Sects.~\ref{subsect:1Dcase} and \ref{subsect:2Dcase}, we used a ray-tracing procedure to accurately compute SEDs. However, in practice, we already know the intensity $\Inu(r=\Rout, \, \Theta, \, \mu, \, \varphi)$ from the DGFEM solution at the outer edge of the envelope. This intensity can theoretically be used to compute the SEDs directly with no further post-processing treatment required. It is, however, not advised to proceed in this way, at least in multi-dimensional cases where the radiation field has a complex geometry, because of the poor accuracy of the local solution on the outer edge due to the poor resolution of the grid ($\Rout \, \Delta \Theta$ is large). This effect becomes even more important if we are interested in studying the inner parts of the disc and hence probing a very small part of the outer shell of the envelope. Consequently a post-processing ray-tracing step is usually needed in order to capture all the features of the disc accurately. The post-processing ray-tracing step requires a good estimate for the emissivity (or source function, see Eq.~\ref{eq:RTE_conservative}) in the whole envelope to compute accurate SEDs (see Appendix \ref{appendix:rayTracer}). The emissivity is implicitly a function of the temperature of the medium (through the black-body emission $B_\nu(T)$ and the radiative equilibrium Eq.~\ref{eq:LTE}), and hence an accurate temperature is needed to compute SEDs with precision.

In order to study the effect of the resolution of the computational grid on the temperature agreement, we computed the $L^2$ norm of the relative residual of the temperature, between our code $T_{\mathrm{DG}}$ and the RADMC-3D code $T_{\mathrm{ref}}$. It is defined as,
\begin{equation}
    L^2 = \left( \frac{ \int\limits_{\mathcal{V}}\left(\frac{T_{\mathrm{DG}} - T_{\mathrm{ref}}}{T_{\mathrm{ref}}} \right)^2 \, r^2 \sin{\Theta} \, dr \, d\Theta}{\int\limits_{\mathcal{V}} r^2 \sin{\Theta} \, dr \, d\Theta} \right)^\frac{1}{2},
\end{equation}
where the integral runs across the volume $\mathcal{V}$ of the envelope. In Fig.~(\ref{fig:error_temperature}), we display the $L^2$ norm versus the total number of elements of our grid (we would like to emphasise that the total number of elements is $N = N_r \times N_\Theta \times N_\mu \times N_\varphi$ and that each element has $3\times2\times3\times3$ nodes). We varied $N$ in such a way as to keep the number of elements along each dimension identical ($N_r = N_\Theta = N_\mu = N_\varphi$). 

We can see that the $L^2$ norm saturates when $N \approx 10^4$, which corresponds to a number of elements along each dimension of ten. After that point, no substantial improvement in the temperature agreement is obtained. The computational grid at this point is still quite coarse, compared to the commonly used methods in the literature such as finite differences \citep{2003A&A...401..405S}, short-characteristics, \citep{2000A&A...360.1187D} or Monte-Carlo \citep{2003CoPhC.150...99W} methods. This is one of the advantages of the DGFEM that can achieve a given error threshold with less resolution, thanks to the form of the intensity inside each element (see Eq.~\ref{eq:Ih_def}). 

In Fig.~\ref{fig:comp_sed_iout_rt}, we compare the SEDs obtained either from the post-processing ray-tracing or directly by using $\Inu(r=\Rout, \, \Theta, \, \mu, \, \varphi)$. We compare these SEDs with the benchmark curves we already presented in  Sects.~\ref{subsect:1Dcase} and \ref{subsect:2Dcase} ($\tau = 100$). While for the spherically symmetric case, they agree quite well, $\Inu(r=\Rout, \, \Theta, \, \mu, \, \varphi)$ fails to give an accurate SED for the 2D disc configuration. This is expected since for this case, the radiation field has a dependence on the $\Theta$ and $\varphi$ variables. We note that the increase in the grid size should make the SED from $\Inu(r=\Rout, \, \Theta, \, \mu, \, \varphi)$ converge to the ray-tracing one, but we could not investigate this hypothesis due to the limits of our current computational resources. In the absence of a finer grid, the post-treatment ray-tracing is thus needed if one wants to compute the emerging fluxes from the disc with precision.


\section{Conclusions} \label{sect:conclusion}

For this study, we applied the DGFEM to the frequency-dependent 2D radiative transfer equation, coupled with the radiative equilibrium equation, inside axis-symmetric circumstellar envelopes. We have shown that it can accurately compute the temperature field and allow for the correct determination of images and SED profiles, via ray-tracing techniques. A desirable feature of the method is the ability to control the order of the numerical scheme via the number of nodes inside each element, meaning that a high-order numerical scheme can simply be achieved by increasing the number of nodes inside each element.

The DGFEM formulation, Eq.~(\ref{eq:DG_FEM}), is particularly adapted to parallelisation and adaptive mesh refinement (AMR) grids \citep[e.g][]{frisken2002}. In this regard, direct access to the residual, Eq.~(\ref{eq:residual}), might provide a robust estimate for the error that could be used as a criterion for the grid refinement \citep[see also][]{2001A&A...380..776R}. We also note that a further 3D generalisation is also possible and straightforward. Together with Monte-Carlo and ray-tracing techniques, the DGFEM provides an additional method for solving the radiative transfer problem, and it could be used in cases where the other methods are expected to be less efficient.


\begin{acknowledgements}
This study has been supported by the Lagrange laboratory of Astrophysics and funded under a 3-year PhD grant from Ecole Doctorale Sciences Fondamentales et Appliquées (EDSFA) of the Université Côte-d'Azur (UCA). The 1D benchmark profiles were computed with the radiative transfer code DUSTY, available at \hyperlink{http://faculty.washington.edu/ivezic/dusty_web}{http://faculty.washington.edu/ivezic/dusty\_web}. The 2D benchmark were computed with RADMC-3D: \hyperlink{https://www.ita.uni-heidelberg.de/~dullemond/software/radmc-3d/}{https://www.ita.uni-heidelberg.de/~dullemond/software/radmc-3d/}
Part of the computations were carried out with the help of OPAL-Meso computing facilities. The authors are grateful to the OPAL infrastructure from Observatoire de la Côte d'Azur (CRIMSON) for providing resources and support. We are grateful to the referee and the editor whose insights helped improve this work. We also want to express our gratitude to the language editor who helped improve the clarity of this paper.
\end{acknowledgements}


\bibliographystyle{aa} 
\bibliography{references} 


\begin{appendix}

\section{Boundary conditions on a spherical enclosed cavity} \label{appendix:bc}

\begin{figure}
    \centering
    \resizebox{\hsize}{!}{\includegraphics{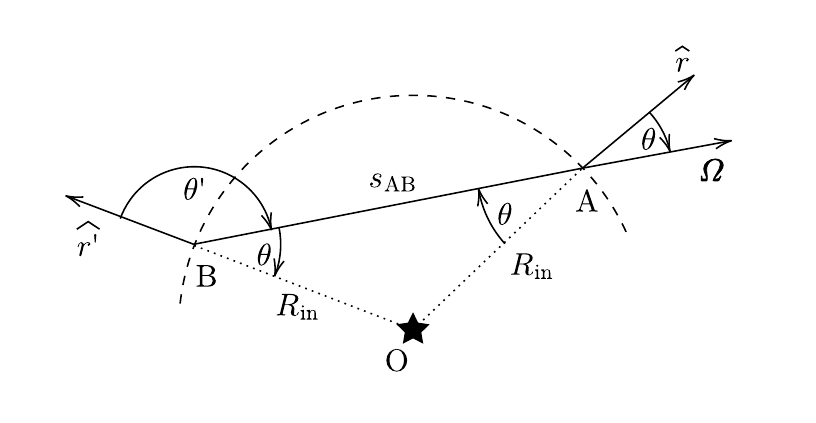}}
    \caption{Geometry of a ray in the inner cavity. Point $B$ is the opposite point of $A$, in the direction $\vecOmega$.}
    \label{fig:inner_boundary}
\end{figure}

The boundary condition in Eq.~(\ref{eq:BC_RT_in}) requires to know, for a given point $A \, (\Rin, \,  \Theta)$ on the inner cavity of radius $\Rin$ and in a direction $\vecOmega \, (\mu, \, \varphi)$ $\left(\vecOmega.\hat{r} > 0\right)$, the coordinates of the opposite point $B$ $\left(\Rin, \, \Theta', \, \mu'=\cos{\theta'}, \, \varphi'\right)$. In the following, the coordinates with a prime superscript are associated with the opposite point $B$. The problem is illustrated in Fig.~\ref{fig:inner_boundary}, where we show the plane containing the point $A$, $B$ and the star. The problem is axis-symmetric around the polar axis and the radiation field does not depend on the azimutal angle $\Phi$ (see Fig.~\ref{fig:coordinates}). Consequently, we can arbitrarily choose the $\Phi$ coordinate of the point $A$ and set it to be $\Phi=\pi/2$. This convention simplifies the computations.

First, we see that $\theta' = \pi - \theta$ because the $OAB$ triangle is isosceles, we then have,
\begin{equation}
    \mu' = -\mu.
\end{equation}
The position $\rv_B$ of the opposite point $B$, is linked to the position $\rv_A$ of the point $A$ through the relation
\begin{equation}
    \rv_B = \rv_A - s_{AB} \, \vecOmega, \label{eq:appendix_1}
\end{equation}
with $s_{AB}$ the distance between the point $A$ and $B$. This length is equal to $s_{AB} = 2 \, \mu \, \Rin$, because it corresponds to the base of the isosceles triangle $OAB$, with $OA = OB = \Rin$. The direction vector, $\vecOmega$, is,
\begin{equation}
\begin{aligned}
    \vecOmega &= \begin{bmatrix} \mu \\ \sqrt{1-\mu^2} \, \cos{\varphi} \\ \sqrt{1-\mu^2} \, \sin{\varphi} \end{bmatrix}_{(\hat{r},\hat{\Theta},\hat{\Phi})}  \\ &= \begin{bmatrix} - \sqrt{1-\mu^2} \, \sin{\varphi} \\ \sin{\Theta} \, \mu  + \cos{\Theta} \, \sqrt{1-\mu^2} \, \cos{\varphi} \\ \cos{\Theta} \, \mu -\sin{\Theta} \, \sqrt{1-\mu^2} \, \cos{\varphi} \end{bmatrix}_{(\hat{x}, \, \hat{y}, \, \hat{z})}. \label{eq:appendix_ov}
\end{aligned}
\end{equation}
Eq.~(\ref{eq:appendix_1}) is then 
\begin{equation}
\begin{aligned}
    \rv_B &= \Rin \begin{bmatrix} \sin{\Theta'} \, \cos{\Phi'} \\ \sin{\Theta'} \, \sin{\Phi'} \\ \cos{\Theta'} \end{bmatrix}_{(\hat{x},\hat{y},\hat{z})} \\  
    &= \Rin \begin{bmatrix} 2 \, \mu \, \sqrt{1-\mu^2} \, \sin{\varphi}  \\ \sin{\Theta} \, \left(1 - 2 \, \mu^2\right) - 2 \, \cos{\Theta} \, \mu \, \sqrt{1-\mu^2} \, \cos{\varphi} \\ \cos{\Theta} \, \left(1 - 2 \, \mu^2\right) + 2 \, \sin{\Theta} \, \mu \, \sqrt{1-\mu^2} \, \cos{\varphi} \end{bmatrix}_{(\hat{x},\hat{y},\hat{z})}. \label{eq:appendix_2}
\end{aligned}
\end{equation}
From the $z$-component of Eq.~(\ref{eq:appendix_2}) we get,
\begin{equation}
    \cos{\Theta'} = \cos{\Theta} \, \left(1 - 2\mu^2\right) + 2 \, \sin{\Theta} \, \mu \, \sqrt{1-\mu^2} \, \cos{\varphi},
\end{equation}
and consequently $\Theta' = \arccos{|\cos{\Theta'}|}$. The absolute value occurs if we consider the planar symmetry with respect to the equatorial plane.

The angle $\varphi'$ at the point $B$ verifies,
\begin{equation}
    \tan{\varphi'} = \frac{\vecOmega.\hat{\Phi'}}{\vecOmega.\hat{\Theta'}}, \label{eq:appendix_phi}
\end{equation}
with,
\begin{equation}
\begin{aligned}
     \hat{\Phi'} &= \begin{bmatrix} - \sin{\Phi'} \\ \cos{\Phi'} \\ 0 \end{bmatrix}_{(\hat{x}, \, \hat{y}, \, \hat{z})} , \, 
     \hat{\Theta'} &= \begin{bmatrix} \cos{\Theta'} \, \cos{\Phi'} \\ \cos{\Theta'} \, \sin{\Phi'} \\ -\sin{\Theta'} \end{bmatrix}_{(\hat{x}, \, \hat{y}, \, \hat{z})}, \\ 
\end{aligned}
\end{equation}
and $\vecOmega$ given by Eq.~(\ref{eq:appendix_ov}). Then, using Eq.~(\ref{eq:appendix_2}) in Eq.~(\ref{eq:appendix_phi}) yields, after tedious calculations,
\begin{equation}
    \tan{\varphi'} = \frac{\sin{\varphi} \, \sin{\Theta}}{\left(1 - 2 \, \mu^2\right) \, \cos{\varphi} \, \sin{\Theta} - 2 \, \mu \, \sqrt{1-\mu^2} \, \cos{\Theta}}. \label{eq:phi'}
\end{equation}
Consequently $\varphi' = \arctan{|u|/v}$ with $u$ and $v$ being the numerator and denominator in Eq.~(\ref{eq:phi'}), respectively. This time, the absolute value is present because the symmetry around the polar axis causes the $\varphi'$ angle to take its values in $[0, \, \pi]$. We note that the previous formula is not valid for the couples of coordinates $\left(\mu, \, \varphi\right)=\left(\sin{(\Theta/2)}, \, 0\right)$ and $\left(\mu, \, \varphi\right) = \left(\cos{(\Theta/2)}, \, \pi\right)$, because it corresponds to have the opposite point $B$ on the poles, where $\hat{\Theta'}$ and $\hat{\Phi'}$ are not defined. These particular cases are in practice not a problem since the radiation field is independent of $\varphi'$ at these points.

\section{DGFEM calculations}\label{appendix:computations}

\begin{figure}
    \centering
    \resizebox{\hsize}{!}{\includegraphics{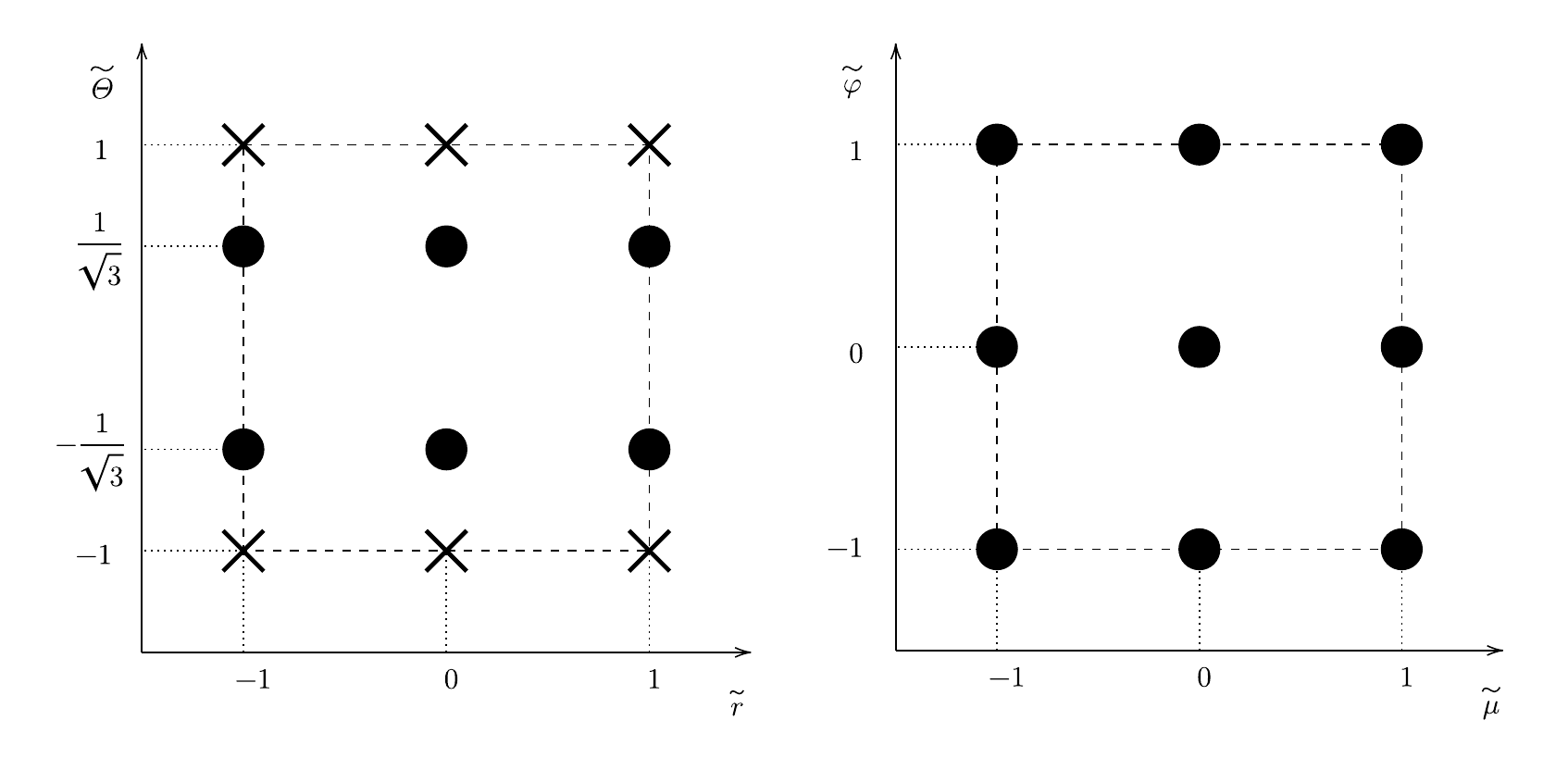}}
    \caption{Example of an element $D^{i,j,k,l}$ (dashed lines) using $n_a=n_c=n_d=3$ and $n_b=2$ points. The left and right panels correspond to 2D slices in the $\left(r,\Theta\right)$ and $\left(\mu,\varphi\right)$ planes, respectively (we display here the local coordinate system given by Eq.~\ref{eq:localcoord}). The black dots correspond to the nodes where the solution is computed while the crosses correspond to the interpolated value of I used to compute the numerical flux $F_h^*$ at the interface, along the $\Theta$ coordinate.}
    \label{fig:element}
\end{figure}

The computation of the terms in Eq.~(\ref{eq:DG_FEM}) requires the choice of a particular quadrature in the cell $D^{i,j,k,l}$, for each coordinate $\left(r, \, \Theta, \, \mu, \, \varphi\right)$. For the $r,\mu,\varphi$ coordinates, we make use of the Gauss-Lobatto quadrature, which has the advantage of having roots at the end points of the cell. This avoids the use of an interpolation formula when computing the numerical flux $\vec{F}_h^{\mathrm{*}}$ at the element edges. The Gauss-Lobatto quadrature can exactly integrate polynomials up to degree $2 \, n - 3$, with $n$ the number of nodes. For the coordinate $\Theta$, we cannot use the Gauss-Lobatto quadrature, at least in the cells that are touching  $\Theta = 0$, because the radiative transfer equation in the spherical coordinates system is not defined at the pole. For this reason and to keep an homogeneous method along $\Theta$, we use a Gauss-Legendre quadrature for this coordinate. The Gauss-Legendre quadrature is more precise and can exactly integrate polynomials up to degree $2n - 1$, but at the expense of an interpolation method required to compute the flux at the cell edges. An example of an element $D^{i,j,k,l}$ with the chosen nodes is shown in Fig.~\ref{fig:element}. 

In the following, all the superscript indexes refer to the element identification while the subscripts denote each node in the considered element. We start with the volume term in Eq.~(\ref{eq:DG_FEM}),
\begin{equation}
\begin{aligned}
    \int \limits_{D^{i,j,k,l}}^{}  \left( \kappa^\mathrm{ext} \, \It_h - \etat  \right) h_{a',b',c',d'} \, d^4\xv \\ 
    =  \frac{\Delta x^{i,j,k,l}}{16} \int \limits_{-1}^{1}  \left( \kappa^\mathrm{ext} \, \It_h^{i,j,k,l} - \etat^{i,j}  \right) \, h_{a',b',c',d'} \, d^4\vec{\Tilde{x}} \\
    = \frac{\Delta x^{i,j,k,l}}{16} \sum\limits_{a,b,c,d} W_{a,b,c,d} \, \left( {\kappa_{a,b}^\mathrm{ext}}^{i,j} \, \It_{a,b,c,d}^{i,j,k,l} - \etat_{a,b}^{i,j}  \right) \, h_{a',b',c',d'}\left(\xt_{a,b,c,d}\right) .
\end{aligned}
\end{equation}
We note that $\Delta x^{i,j,k,l} = \Delta r^i \, \Delta \, \Theta^j \, \Delta \mu^k \, \Delta \varphi^l $ is the 4D volume of the element $D^{i,j,k,l}$. For integration, we rather use the local coordinates $(\Tilde{r}, \, \Tilde{\Theta}, \, \Tilde{\mu}, \, \Tilde{\varphi})$, defined as (e.g for the r coordinate),
\begin{equation}
    \Tilde{r} = \frac{2}{\Delta r^i} \left(r - r^{i+1/2} \right), \label{eq:localcoord}
\end{equation}
with $\Delta r^i$, the element width along the coordinate r and $r^{i+1/2}$, the radial coordinate of the centre of the element. The same expression holds for the other coordinates. The quantities $W_{a,b,c,d} = W_r(\rt_a) \, W_\Theta(\Tt_b) \, W_\mu(\mut_c) \, W_\varphi(\phit_d)$ are the weights associated with the different quadrature in each direction. Finally, $h_{a',b',c',d'}\left(\vec{\Tilde{x}}_{a,b,c,d}\right)$ is the 4D Lagrange polynomials, defined in Eq.~(\ref{eq:lagrange_poly}), evaluated at the node $\xt_{a,b,c,d}= (\Tilde{r}_a, \, \Tilde{\Theta}_b, \, \Tilde{\mu}_c, \, \Tilde{\varphi}_d)$. By definition of the Lagrange polynomials, we have,
\begin{equation}
    h_{a',b',c',d'}\left(\rt_a, \, \Tt_b, \, \mut_c, \, \phit_d\right) = \delta_{a',a} \, \delta_{b',b} \, \delta_{c',c} \, \delta_{d',d},
\end{equation}
with $\delta_{a',a}$ the usual delta Kronecker. The other volume term in Eq.~(\ref{eq:DG_FEM}) is expressed as,
\begin{equation}
\begin{aligned}
    \int \limits_{D^{i,j,k,l}}^{} \vec{F}_h.\nabla_{\xv} h_{a',b',c',d'}  \, d^4\xv  \\ 
    = \int \limits_{D^{i,j,k,l}} a_r \, \Ih \,  \partial_r h_{a',b',c',d'}  \, d^4\xv 
    + \int \limits_{D^{i,j,k,l}} a_\Theta \, \Ih \, \partial_\Theta h_{a',b',c',d'}   \, d^4\xv \\
    + \int \limits_{D^{i,j,k,l}} a_\mu \, \Ih \, \partial_\mu h_{a',b',c',d'}   \, d^4\xv
    + \int \limits_{D^{i,j,k,l}} a_\varphi \, \Ih \, \partial_\varphi h_{a',b',c',d'}   \, d^4\xv \\
    = \frac{\Delta x^{i,j,k,l}}{8}  \sum\limits_{a,b,c,d} W_{a,b,c,d} \, \It_{a,b,c,d}^{i,j,k,l} \left( \frac{{a_{\rt}}_{a,b,c,d}^{i,j,k,l} \, \partial_{\rt} h_{a',b',c',d'}|_{\xt_{a,b,c,d}}}{\Delta r^i} \right. \\
     \left. + \frac{{a_{\Tt}}_{a,b,c,d}^{i,j,k,l} \, \partial_{\Tt} h_{a',b',c',d'}|_{\xt_{a,b,c,d}}}{\Delta \Theta^j} + \frac{{a_{\mut}}_{a,b,c,d}^{i,j,k,l} \, \partial_{\mut} h_{a',b',c',d'}|_{\xt_{a,b,c,d}}}{\Delta \mu^k} \right. \\ \left. + \frac{{a_{\phit}}_{a,b,c,d}^{i,j,k,l} \, \partial_{\phit} h_{a',b',c',d'}|_{\xt_{a,b,c,d}}}{\Delta \varphi^l} \right). \label{eq:appendix_deriv}
\end{aligned}
\end{equation}
In Eq.~(\ref{eq:appendix_deriv}), we used the definition of the flux Eq.~(\ref{eq:flux_def}). The quantity $\partial_{\rt} h_{a',b',c',d'}|_{\xv_{a,b,c,d}}$ is the partial derivative of the Lagrange polynomial, with respect to the coordinate $r$, evaluated at the node $\xt_{a,b,c,d}$ (with similar definitions for the other coordinates).

The last term to evaluate is the surface integral (first term in Eq.~\ref{eq:DG_FEM}). The 4D element is delimited by $2 \times 4 = 8$ surfaces. We give here the derivation for the surface integrals normal to the coordinate $r$ and the other terms will follow by substitution of the indices. We have $\shat = \pm \rhat$ for the radial right and left surfaces ($\rt=\pm1$), respectively,   
\begin{equation}
\begin{aligned}
    \oint \limits_{\partial D^{i,j,k,l}} \left[F_r^{\mathrm{*}} h_{a',b',c',d'} \right]_{\rt=-1}^{\rt=1}  \, d\Theta \, d\mu \, d\varphi \\
    =  \frac{\Delta x^{i,j,k,l}}{8 \, \Delta r^i} W_{b',c',d'} \left[F_{\rt}^{\mathrm{*}}\left(\rt,\Tt_{b'},\mut_{c'},\phit_{d'}\right) \, h_{a'}\left(\rt\right) \right]_{\rt=-1}^{\rt=1}. \label{eq:surface_r}
\end{aligned}
\end{equation}
At the cell boundaries, we use the upwind numerical flux, for example, at the right edge,
\begin{equation}
    \begin{aligned}
        F_{\rt}^{\mathrm{*}}\left(\rt = 1,\Tt_{b'},\mut_{c'},\phit_{d'}\right) = 
        \\ \max{\left\lbrace a_{\rt}^{i,j,k,l}\left(1,\Tt_{b'},\mut_{c'},\phit_{d'}\right) ,0\right\rbrace} \, \Ih^{i,j,k,l}\left(1,\Tt_{b'},\mut_{c'},\phit_{d'}\right) \\ + \min{\left\lbrace a_{\rt}^{i,j,k,l}\left(1,\Tt_{b'},\mut_{c'},\phit_{d'}\right) ,0\right\rbrace} \, \Ih^{i+1,j,k,l}\left(-1,\Tt_{b'},\mut_{c'},\phit_{d'}\right) \, .
    \end{aligned}
\end{equation}
The use of the Gauss-Lobatto quadrature further simplify the computation because the quadrature nodes include the end-points $\rt = \pm1$. The numerical flux can then be expressed as
\begin{equation}
\begin{aligned}
       \left[F_{\rt}^{\mathrm{*}}\left(\rt,\Tt_{b'},\mut_{c'},\phit_{d'}\right) h_{a'}\left(\rt\right) \right]_{\rt=-1}^{\rt=1} = \\ 
       \delta_{a',n_a-1} \max{\left\lbrace {a_{\rt}}_{n_a-1,b',c',d'}^{i,j,k,l}, 0\right\rbrace} \, \It_{n_a-1,b',c',d'}^{i,j,k,l} \\
     + \delta_{a',n_a-1} \min{\left\lbrace {a_{\rt}}_{n_a-1,b',c',d'}^{i,j,k,l}, 0\right\rbrace} \, \It_{0,b',c',d'}^{i+1,j,k,l} \\
     - \delta_{a',0} \max{\left\lbrace {a_{\rt}}_{0,b',c',d'}^{i,j,k,l}, 0\right\rbrace} \, \It_{n_a-1,b',c',d'}^{i-1,j,k,l} \\
      - \delta_{a',0} \min{\left\lbrace {a_{\rt}}_{0,b',c',d'}^{i,j,k,l}, 0\right\rbrace} \, \It_{0,b',c',d'}^{i,j,k,l}.  \label{eq:flux_integral_r}
\end{aligned}
\end{equation}
The same form holds for the surface integrals normal to the $\mu$ and $\varphi$ coordinates. We however note that we have $a_\mu \geq 0, \, a_\varphi \leq 0 \, \forall \, \xv \in D$, so there are no terms proportional to $\It^{i,j,k+1,l}$ and $\It^{i,j,k,l-1}$. For the flux computation normal to the $\Theta$ coordinate, we do not directly have the value of the solution at the element interface (see Fig.~\ref{fig:element}), we need to interpolate the solution with the help of Eq.~(\ref{eq:Ih_def}),
\begin{equation}
    \begin{aligned}
           F_{\Tt}^{\mathrm{*}}\left(\rt_{a'},\Tt = 1,\mut_{c'},\phit_{d'}\right) = \\ \max{\left\lbrace a_{\Tt}^{i,j,k,l}\left(\rt_{a'}, 1,\mut_{c'},\phit_{d'}\right) ,0\right\rbrace}  \sum\limits_b  \It_{a',b,c',d'}^{i,j,k,l} h_b\left(1\right) \\ + \min{\left\lbrace a_{\Tt}^{i,j,k,l}\left(\rt_{a'}, 1,\mut_{c'},\phit_{d'}\right) ,0\right\rbrace}  \sum\limits_b  \It_{a',b,c',d'}^{i,j+1,k,l} h_b\left(-1\right).
    \end{aligned}
\end{equation}

All the terms in this section can be put in the form of the system of equations, given by Eq.~(\ref{eq:system_iteration}),
\begin{equation}
    \mathcal{A}^{i,j,k,l} \, \Ih^{i,j,k,l} = \vec{b}^{i,j,k,l}. \label{eq:system_appendix}
\end{equation}
In Eq.~(\ref{eq:system_appendix}), we replaced the RHS by $\vec{b}^{i,j,k,l}$ because we do not need to formally write the non-diagonal matrices $\mathcal{A}^{i\pm 1,j\pm1,k-1,l+1}$. To assemble $\mathcal{A}^{i,j,k,l}$, we make use of the global index $\alpha = a \, n_b n_c n_d + b \, n_c n_d + c \, n_d + d $. The elements of $\mathcal{A}^{i,j,k,l}$ and $\vec{b}^{i,j,k,l}$ are then,
\begin{equation}
\begin{split}
\mathcal{A}_{\alpha'\alpha}^{i,j,k,l} = 
W_{a,b,c,d} \, {\kappa_{a,b}^{\mathrm{ext}}}^{i,j} \delta_{a',a} \, \delta_{b',b} \, \delta_{c',c} \, \delta_{d',d} \\
+ \frac{2}{\Delta r^i} \delta_{b',b}\delta_{c',c}\delta_{d',d} W_{b,c,d} \left(  \max{\left\lbrace {a_{\rt}}_{a,b,c,d}^{i,j,k,l},0\right\rbrace} \delta_{a',n_a-1} \delta_{a,n_a-1} \right. \\ \left. - \min{\left\lbrace {a_{\rt}}_{a,b,c,d}^{i,j,k,l},0\right\rbrace} \delta_{a',0}\delta_{a,0} - W_a \, {a_{\rt}}_{a,b,c,d}^{i,j,k,l} \, \partial_{\rt} h_{a'}|_{\rt_{a}} \right) \\ 
+ \frac{2}{\Delta \Theta^j} \delta_{a',a}\delta_{c',c}\delta_{d',d} W_{a,c,d} \left(  \max{\left\lbrace {a_{\Tt}}_{a,c,d}^{i,j,k,l}|_{\Tt=1},0\right\rbrace} h_{b'}|_{\Tt=1} h_{b}|_{\Tt=1} \right. \\ \left.
- \min{\left\lbrace {a_{\Tt}}_{a,c,d}^{i,j,k,l}|_{\Tt=-1},0\right\rbrace} h_{b'}|_{\Tt=-1} h_{b}|_{\Tt=-1} - W_b \, {a_{\Tt}}_{a,b,c,d}^{i,j,k,l} \, \partial_{\Tt} h_{b'}|_{\Tt_{b}} \right) \\ 
+ \frac{2}{\Delta \mu^k} \delta_{a',a}\delta_{b',b}\delta_{d',d} W_{a,b,d} {a_{\mut}}_{a,b,c,d}^{i,j,k,l} \left( \delta_{c',n_c-1} \delta_{c,n_c-1} - W_c \partial_{\mut} h_{c'}|_{\mut_{c}} \right) \\ 
- \frac{2}{\Delta \varphi^l} \delta_{a',a}\delta_{b',b}\delta_{c',c} W_{a,b,c} {a_{\phit}}_{a,b,c,d}^{i,j,k,l} \left( \delta_{d',0}\delta_{d,0} + W_d \partial_{\phit} h_{d'}|_{\phit_d} \right).
\end{split}
\end{equation}
\begin{equation}
    \begin{split}
        b_{\alpha'}^{i,j,k,l} = W_{a',b',c',d'} \, \etat_{a',b'}^{i,j} \\
        + \frac{2}{\Delta r^i} W_{b',c',d'}\left( \max{\left\lbrace {a_{\rt}}_{n_a-1,b',c',d'}^{i-1,j,k,l},0\right\rbrace} \It_{n_a-1,b',c',d'}^{i-1,j,k,l}\delta_{a',0} \right. \\ \left.
        - \min{\left\lbrace {a_{\rt}}_{0,b',c',d'}^{i+1,j,k,l},0\right\rbrace} \It_{0,b',c',d'}^{i+1,j,k,l}\delta_{a',n_a-1}  \right) \\
        + \frac{2}{\Delta \Theta^j}  W_{a',c',d'}  \left( \max{\left\lbrace {a_{\Tt}}_{a',c',d'}^{i,j-1,k,l}|_{\Tt = 1},0\right\rbrace}h_{b'}|_{\Tt = -1}  
         \sum\limits_{b} \It_{a',b,c',d'}^{i,j-1,k,l} h_{b}|_{\Tt = 1} \right.\\ \left. - \min{\left\lbrace {a_{\Tt}}_{a',c',d'}^{i,j+1,k,l}|_{\Tt = -1},0\right\rbrace}h_{b'}|_{\Tt = 1} 
         \sum\limits_{b} \It_{a',b,c',d'}^{i,j+1,k,l} h_{b}|_{\Tt = -1} \right)  \\
        + \frac{2}{\Delta \mu^k}  W_{a',b',d'} {a_{\mut}}_{a',b',n_c-1,d'}^{i,j,k-1,l}  \It_{a',b',n_c-1,d'}^{i,j,k-1,l}\delta_{c',0} \\
        - \frac{2}{\Delta \varphi^l} W_{a',b',c'} {a_{\phit}}_{a',b',c',0}^{i,j,k,l+1} \It_{a',b',c',0}^{i,j,k,l+1}\delta_{d',n_d-1}.
    \end{split}
\end{equation}

\section{Ray-tracing module}\label{appendix:rayTracer}

\begin{figure}
    \centering
    \resizebox{\hsize}{!}{\includegraphics{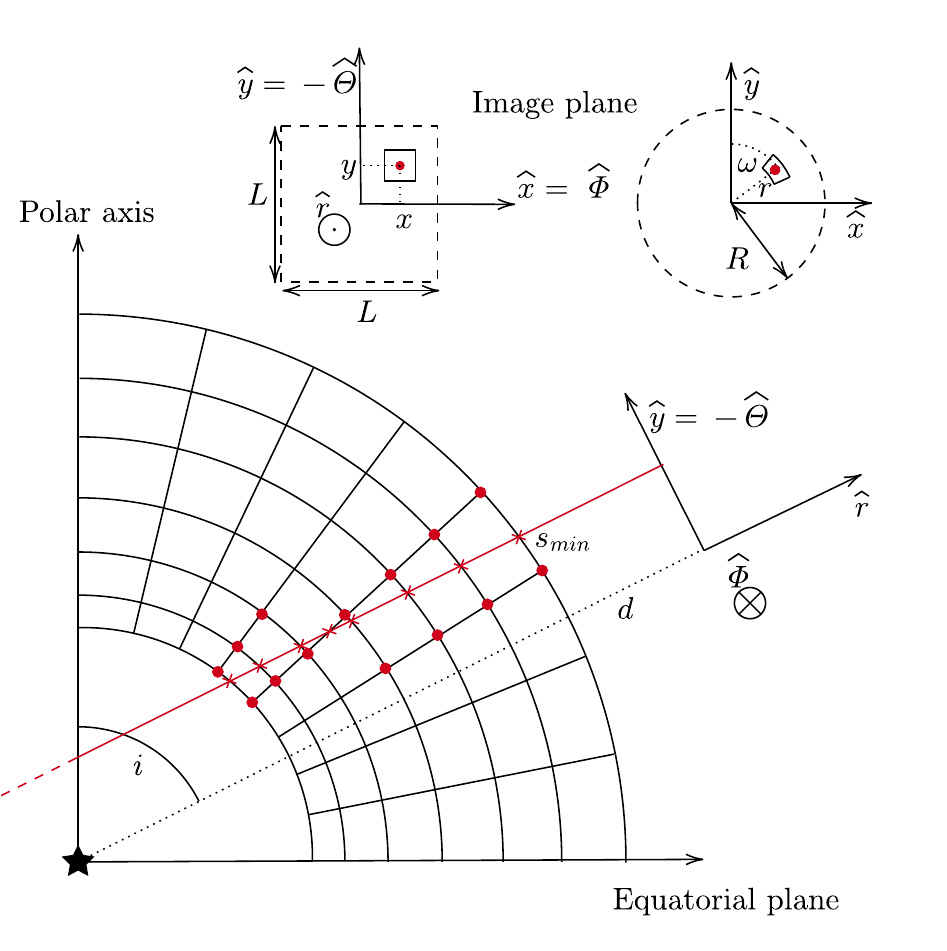}}
    \caption{Example of a ray (red line) normal to the image plane, crossing the spherical grid. In this example, we display a square image of size $L \times L$ and a circular image of radius $R$. The red crosses represent the intersections between the ray and the grid. The values of $\Knuext$ and $S_\nu$ at these intersections are linearly interpolated from the grid adjacent values (red dots).}
    \label{fig:ray_tracer}
\end{figure}

The SEDs and intensity maps from the DGFEM code, shown in Sect~\ref{sect:num_tests}, were computed with the help of the ray-tracing routine we present here. This procedure is quite generally used in the literature, for all type of codes, and is not a limitation of the DGFEM method itself \citep[see e.g][Sect 2.2.3]{2009A&A...498..967P}

For the SED, we need to estimate the total flux $f_\nu^\mathrm{obs}$ that an observer receive from the object, situated at a distance $d \gg \Rout$ and doing an angle $i$ with the polar axis (see Fig.~\ref{fig:ray_tracer}). Because we decoupled the stellar from the envelope radiation (see Eq.~\ref{eq:RTE2}), the total flux is made of the stellar and envelope total flux,
\begin{equation}
    f_\nu^\mathrm{obs}(i,d) = f_\nu^\mathrm{obs,\star} + f_\nu^\mathrm{obs,env}. \label{eq:sum_flux}
\end{equation}
If we assume the star to be an unresolved black-body point source, the stellar flux at distance $d$ is the flux of the star at the stellar surface attenuated by the circumstellar matter present in the direction of the line of sight (black dotted line in Fig.~\ref{fig:ray_tracer}), and with the dilution factor $(\Rstar/d)^2$,
\begin{equation}
\begin{aligned}
    f_\nu^\mathrm{obs,\star}(i,d) &= \pi \, \left(\frac{\Rstar}{d}\right)^2 \, \Bnu(\Tstar) \, \exp{\left\lbrace-\tau(i)\right\rbrace} \label{eq:fobs_star} \\
    \text{with} \quad \tau(i) &= \int_{\Rin}^{\Rout} \Knuext(r,i) \, dr
\end{aligned}
\end{equation}

To compute the envelope flux and intensity maps, we define an image plane $\left(\hat{x}, \, \hat{y}\right)$, at a distance $d \gg \Rout$ from the star and tilted with an angle $i$ with respect to the polar axis. The $x$ and $y$ axes are oriented with the help of the spherical coordinates system $(\hat{r}, \, \hat{\Theta}, \, \hat{\Phi})$. In this plane, we can construct images of any geometry but we only consider the special cases of a square image of width $L$ and a circular image of radius $R$. Since $d \gg \Rout$, the flux inside the image can formally be written,
\begin{equation}
\begin{aligned}
     f_\nu^\mathrm{obs,env}(i) &= \int \limits_{-\frac{L}{2}}^{\frac{L}{2}} \int \limits_{-\frac{L}{2}}^{\frac{L}{2}} \Inuenv(x,y,\hat{r}) \, \frac{dx \, dy}{d^2}, \\
      \text{or} \quad f_\nu^\mathrm{obs,env}(i) &= \int \limits_{0}^{2\pi} \int \limits_{0}^{R} \Inuenv(r,\omega,\hat{r}) ~ \frac{r \, d\omega \, dr}{d^2}.
\end{aligned}
\end{equation}
In practice, images are made of a collection of pixels in which we evaluate the emerging specific intensity at the pixel centre. A square (circular) image, is divided into $N\times N$ ($N_r\times N_\omega$) pixels and the flux in the image can then be rewritten,
\begin{equation}
\begin{aligned}
    f_\nu^\mathrm{obs,env}(i) &\approx \sum \limits_{i=0}^{N-1} \sum \limits_{j=0}^{N-1} \Inuenv(x_i,y_j,\hat{r}) ~ \frac{\Delta x_i \, \Delta y_i}{d^2}, \\
    \text{or} \quad f_\nu^\mathrm{obs,env}(i) &\approx \sum \limits_{i=0}^{N_r} \sum \limits_{j=0}^{N_\omega} \Inuenv(r_i,\omega_j,\hat{r}) ~ \frac{r_i \, \Delta \omega_j \, \Delta r_i}{d^2},
\end{aligned}
\end{equation}
with $\Delta x_i \, \Delta y_i$ ($r_i \, \Delta \omega_j \, \Delta r_i$) the pixel size of the square (circular) image. We note that the circular image is particularly well-suited for the computation of $f_\nu^\mathrm{obs,env}(i)$ since we can easily increase the number of pixels in the centre of the image in order to resolve the disc inner parts.

The emerging specific intensity crossing each pixel centre $\rv_0 = x_i \, \hat{x} + y_j \, \hat{y} + d \, \hat{r}$ ($\rv_0 = r_i \, \sin{\omega_j} \, \hat{x} + r_i \, \cos{\omega_j} \, \hat{y} + d \, \hat{r}$ for a circular image), along the ray normal to the image plane (red line in Fig.~\ref{fig:ray_tracer}) is computed by integration of the emissivity along the ray,
\begin{equation}
\begin{aligned}
         \Inuenv(\rv_0,\rhat) &= \int \limits_{s_{min}}^{s_{max}} \eta_\nu \exp{\left\lbrace-\tau_\nu(s)\right\rbrace} \, ds, \\
         \text{with} \quad  \tau_\nu(s) &= \int\limits_{s_{min}}^{s} \Knuext \, ds'.
\end{aligned}
\end{equation}
We define $s$ as to be the distance from the pixel centre $\rv_0$ to a given point along the ray. The quantities $\eta_\nu$ and $\Knuext$ are the emissivity and the extinction coefficient, respectively, as defined in Eq.~(\ref{eq:RTE1}). The points $s_{min}$ and $s_{max}$ correspond to the two intersections of the ray with the sphere of radius $\Rout$. 

In practice, $\eta_\nu$ and $\Knuext$ are defined on a discrete grid and the previous integral can be rewritten as,
\begin{equation}
    \Inuenv(\rv_0,\rhat) = \sum\limits_{i=0}^{n-2} \int \limits_{s_i}^{s_{i+1}} \eta_\nu \, \exp{\left\lbrace-\tau_\nu(s)\right\rbrace} \, ds.
\end{equation}
The $\left\lbrace s_i\right\rbrace$ are the $n$ intersections between the ray and the grid (red crosses in Fig.~\ref{fig:ray_tracer}), with $s_0=s_{min}$ and $s_{n-1}=s_{max}$. The coordinates of all intersections can be computed, in the Cartesian coordinate system. We can express $\Delta \Inu^{i}$, the contribution to  the total intensity $\Inuenv(x, \, y, \, \rhat)$ of each portion between two consecutive intersections as,
\begin{equation}
\begin{aligned}
     \Inuenv(\rv_0, \, \rhat) &= \sum\limits_{i=0}^{n-2} \exp{\left(-\tau_\nu^i\right)} \, \Delta \Inu^i, \\  
     \text{with} \ \Delta \Inu^i &=  \int \limits_{s_i}^{s_{i+1}} \eta_\nu \, \exp{\left(- \int\limits_{s_{i}}^{s} \Knuext \, ds' \right)} \, ds, \\
     \text{and} \ \tau_\nu^i &= \int\limits_{s_{min}}^{s_i}  \Knuext \, ds'.
\end{aligned}
\end{equation}
Following \citet{1986JQSRT..35..431O}, we assume that $\eta_\nu$ and $\Knuext$ are linear functions between two consecutive intersections. Each contribution $\Delta \Inu^{i}$ is given by,
\begin{equation}
\begin{aligned}
    \Delta \Inu^i &\approx \left( 1 - \exp{\left\lbrace - \Delta \tau_\nu^i \right\rbrace} - \beta \right) S_\nu(s_i) + \beta \, S_\nu(s_{i+1}) \, , \\
    \text{with} \ \beta &= \frac{\Delta \tau_\nu^i - 1 + \exp{\left\lbrace - \Delta \tau_\nu^i \right\rbrace}}{\Delta \tau_\nu^i} , \\ 
    \Delta \tau_\nu^i &= \frac{\Knuext(s_i)+\Knuext(s_{i+1})}{2}\left( s_{i+1} - s_i\right),
    \end{aligned} \label{eq:DI}
\end{equation}
and $S_\nu = \eta_\nu / \Knuext$. The values of $S_\nu$ and $\Knuext$ at the intersections $s_i$ and $s_{i+1}$ are estimated by linear interpolation from the grid-adjacent values (red dots in Fig.~\ref{fig:ray_tracer}). The optical depth, $\tau_\nu^i$, can be computed recursively,
\begin{equation}
    \tau_\nu^{i+1} = \tau_\nu^{i} + \int\limits_{s_{i}}^{s_{i+1}} \Knuext \, ds' = \tau_\nu^{i} + \Delta \tau_\nu^i \, ,
\end{equation}
with $\tau_\nu^0 = 0$ and $\Delta \tau_\nu^i$ defined in Eq.~(\ref{eq:DI}).
\end{appendix}

\end{document}